\PassOptionsToPackage{unicode}{hyperref}
\PassOptionsToPackage{hyphens}{url}
\documentclass[
]{article}
\usepackage{xcolor}
\usepackage{amsmath,amssymb}
\usepackage{iftex}
\ifPDFTeX
  \usepackage[T1]{fontenc}
  \usepackage[utf8]{inputenc}
  \usepackage{textcomp}
\else
  \usepackage{unicode-math}
  \defaultfontfeatures{Scale=MatchLowercase}
  \defaultfontfeatures[\rmfamily]{Ligatures=TeX,Scale=1}
\fi
\usepackage{lmodern}
\ifPDFTeX\else
\fi
\IfFileExists{upquote.sty}{\usepackage{upquote}}{}
\IfFileExists{microtype.sty}{
  \usepackage[]{microtype}
  \UseMicrotypeSet[protrusion]{basicmath}
}{}
\makeatletter
\@ifundefined{KOMAClassName}{
  \IfFileExists{parskip.sty}{%
    \usepackage{parskip}
  }{
    \setlength{\parindent}{0pt}
    \setlength{\parskip}{6pt plus 2pt minus 1pt}}
}{
  \KOMAoptions{parskip=half}}
\makeatother
\usepackage{longtable,booktabs,array}

\usepackage{multirow}
\usepackage{calc}
\usepackage{etoolbox}
\makeatletter
\patchcmd\longtable{\par}{\if@noskipsec\mbox{}\fi\par}{}{}
\makeatother
\IfFileExists{footnotehyper.sty}{\usepackage{footnotehyper}}{\usepackage{footnote}}
\makesavenoteenv{longtable}
\usepackage{graphicx}
\makeatletter
\newsavebox\pandoc@box
\newcommand*\pandocbounded[1]{
  \sbox\pandoc@box{#1}%
  \Gscale@div\@tempa{\textheight}{\dimexpr\ht\pandoc@box+\dp\pandoc@box\relax}%
  \Gscale@div\@tempb{\linewidth}{\wd\pandoc@box}%
  \ifdim\@tempb\p@<\@tempa\p@\let\@tempa\@tempb\fi
  \ifdim\@tempa\p@<\p@\scalebox{\@tempa}{\usebox\pandoc@box}%
  \else\usebox{\pandoc@box}%
  \fi%
}
\def\fps@figure{htbp}
\makeatother
\ifLuaTeX
  \usepackage{luacolor}
  \usepackage[soul]{lua-ul}
\else
  \usepackage{soul}
\fi

\makeatletter
 \let\@cite@ofmt\@firstofone
 \def\@biblabel#1{}
 \def\@cite#1#2{{#1\if@tempswa , #2\fi}}
\makeatother
\newlength{\cslhangindent}
\newlength{\csllabelwidth}
\usepackage{calc}

\usepackage{bookmark}
\IfFileExists{xurl.sty}{\usepackage{xurl}}{}
\hypersetup{
  hidelinks,
  pdfcreator={LaTeX via pandoc}}

\usepackage[letterpaper,margin=1in]{geometry}
\usepackage{fancyhdr}
\fancypagestyle{plain}{\fancyhf{}\fancyfoot[C]{\thepage}}

\usepackage[most]{tcolorbox}
\usepackage{capt-of}
\usepackage{enumitem}
\newlist{qlist}{itemize}{2}
\setlist[qlist,1]{label=\textendash, leftmargin=1.3em, topsep=3pt, itemsep=3pt, parsep=0pt}
\setlist[qlist,2]{label=\textperiodcentered, leftmargin=1.2em, topsep=3pt, itemsep=1pt, parsep=0pt}
\newcommand{\qcat}[1]{\smallskip\noindent\textit{\textbf{#1}}\par\nobreak}
\newtcolorbox{chapbox}{breakable, enhanced, colback=black!3, colframe=black!45,
  boxrule=0.6pt, left=10pt, right=10pt, top=9pt, bottom=9pt}

\author{}
\date{}

\begin{document}

\thispagestyle{plain}

\begin{center} {\large\scshape Chapter 8}\\[0.9em] {\LARGE\bfseries The uses (and misuses) of Earth Observation data\\[0.25em] for weather and vegetation analysis\par}
\bigskip
  \begin{tabular}{c}
    Elinor Benami, Virginia Tech \textperiodcentered{} ORCID \href{https://orcid.org/0000-0002-8586-1314}{0000-0002-8586-1314}\\[0.35em]
    Jeffrey D. Michler, University of Arizona \textperiodcentered{} ORCID \href{https://orcid.org/0000-0001-7778-8703}{0000-0001-7778-8703}\\[0.35em]
    Anna Josephson, University of Arizona \textperiodcentered{} ORCID \href{https://orcid.org/0000-0001-6739-7442}{0000-0001-6739-7442}\\[0.35em]
    Michael Cecil, University of Maryland \textperiodcentered{} ORCID \href{https://orcid.org/0000-0003-2011-8607}{0000-0003-2011-8607}\\[0.35em]
    Gina Maskell, Potsdam Institute for Climate Impact Research (PIK) \textperiodcentered{} ORCID \href{https://orcid.org/0000-0001-7903-2512}{0000-0001-7903-2512}\\
  \end{tabular}\\[1.8em]
  {\small\itshape Forthcoming chapter from \emph{Geospatial Impact Evaluation in Practice},\\
   eds. Ariel Ben-Yishay and Kunwar Singh}
\end{center}

\vspace{1.5em}

\section*{Executive summary}
Integrating gridded Earth observation and weather data into impact evaluations holds great promise. These data allow researchers to capture environmental context, external shocks, and intervention outcomes (e.g., land cover change and agricultural production) that surveys might miss due to spatial or temporal data collection constraints. However, with great power comes great responsibility: The growing ease with which researchers can extract and analyze time series from these datasets can obscure complex geospatial and measurement issues affecting the magnitude, direction, and interpretation of impact estimates. This chapter highlights common challenges associated with the use of weather, vegetation, and extreme event data in the context of geospatial impact evaluation, while providing practical guidance and resources to help researchers judiciously use and avoid misusing these datasets.

\textbf{Keywords:} remote sensing, agriculture, Earth observation, measurement error, data generating process, impact evaluation, weather and environmental data

\textbf{Learning objectives:}

\begin{enumerate} \def\labelenumi{\arabic{enumi}.}
\item Distinguish between key types of Earth observation (EO) and weather data products, including interpolated station data, EO-derived products, merged datasets, and reanalysis products, while evaluating their respective strengths and limitations.
\item Identify common sources of measurement error in EO data and apply strategies to improve the relevance and appropriateness of selected measures.
\item Evaluate the suitability of EO products by aligning product characteristics with research design and analytical needs.
\item Design more robust analyses by assessing the implications of data selection and applying better practices for transparency and reproducibility in EO-integrated evaluations.
\end{enumerate}

\clearpage
\section{Introduction}

Impact evaluations are increasingly using gridded weather and Earth observation (EO) data to capture environmental context, characterize abnormal conditions (shocks), and measure outcomes. Such data can be especially valuable when ground-based information is sparse, delayed, or collected at scales that do not align well with the intervention or outcome of interest. Yet the ease of access to ready-made products can obscure the design choices embedded in their construction. The same underlying phenomenon can look different depending on how raw observations are collected and transformed into analysis-ready datasets. As a result, products with seemingly similar labels such as ``rainfall,'' ``temperature,'' or ``vegetation'' may capture different constructs than intended, potentially affecting the magnitude, direction, and interpretation of estimated effects.

This chapter provides practical guidance on how to select and use various weather and EO-derived data products appropriately within impact evaluation research designs. We aim to inform economists and quantitative social scientists about the opportunities and pitfalls these data present within econometric research designs used to estimate causal effects, while also helping those with remote sensing or geospatial science backgrounds understand how EO data can introduce both systematic and nonsystematic sources of bias into applied, policy-relevant research. More broadly, our goal is to bridge these scholarly communities around a shared concern: how measurement error in gridded weather and EO-based data can bias impact evaluation estimates and, by extension, the policy guidance derived from them.

\begin{figure}[htbp]
  \centering
  \includegraphics[width=\textwidth,height=0.75\textheight,keepaspectratio]{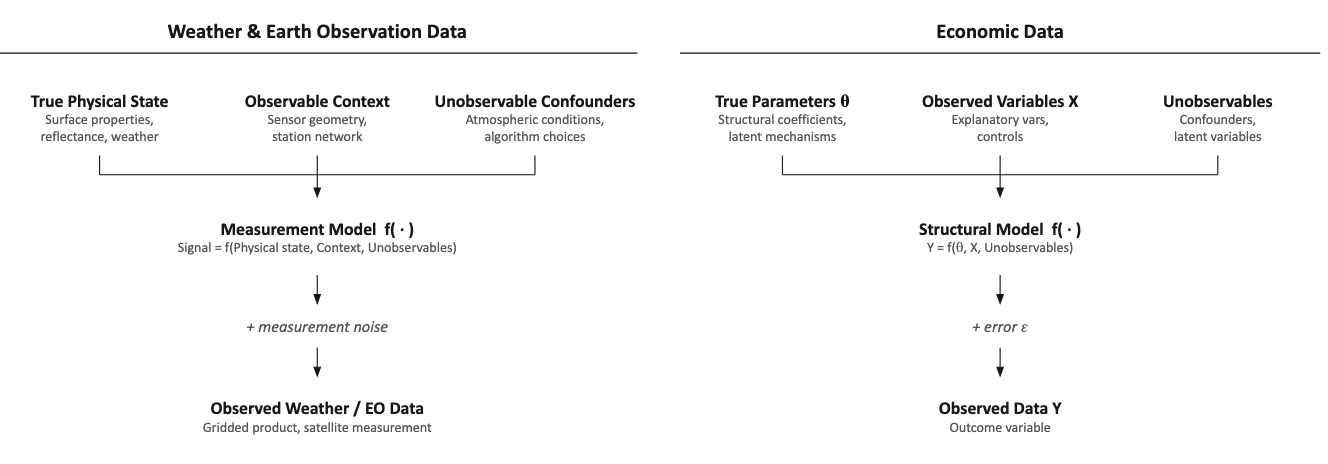}
  \caption{Both the weather/EO DGP and the economic DGP describe how observable data are generated from a true underlying state via a noisy transformation. The DGPs are structurally parallel but serve different inferential purposes: The weather and EO data side primarily involves a measurement problem: how faithfully observed data represent an underlying physical quantity. By contrast, the economic DGP involves an identification problem, where the challenge is credibly recovering latent causal relationships from observed data and modeling assumptions.}
  \label{fig:8-1}
\end{figure}

Central to this chapter is the concept of a data generating process (DGP) for gridded weather and EO data: the set of physical conditions, sensor characteristics, modeling assumptions, and processing choices that jointly determine what is measured and recorded in an analysis-ready dataset. Just as credible causal inference requires understanding the structural relationships and sources of variation that generate the data used in analysis (Abadie et al. 2020), making the EO data generating process explicit helps clarify when weather and EO-derived variables reflect the construct of interest and when they instead capture artifacts or confounding variation introduced by the measurement process itself. Figure~\ref{fig:8-1} illustrates the parallels between the EO data generating process and the economic DGP.

When gridded weather or EO outputs are used as inputs in economic models, these nested DGPs create nested error structures that can compound measurement problems (Hausman 2001; Schennach 2016). Differences arising from factors such as atmospheric conditions, the assumptions embedded in station interpolation techniques, or sensor acquisition angle may introduce classical measurement error, which reduces estimation efficiency, or non-classical measurement error, which biases estimates in ways that depend on the research context and therefore cannot be assumed away. Ultimately, what is at stake is the credibility of inference in geospatial impact evaluations (GIEs) that rely on these data.

The sections that follow translate these conceptual foundations into actionable guidance. We begin with weather data, covering commonly used measures such as precipitation and temperature alongside important but less frequently used variables such as evapotranspiration and vapor pressure deficit. We then turn to EO-derived vegetation data, including how spectral signals are converted into indices and where errors or misinterpretation can arise. This is followed by a discussion of the challenges involved in detecting extreme events and defining weather shocks. The chapter concludes with strategies for addressing measurement issues and practices that support rigorous and transparent GIEs.

\section{Weather and atmospheric data}

The goal of many remotely sensed EO data products is to record observable phenomena at the Earth's surface and in the atmosphere from a distance and convert them into estimates of environmental conditions (EO data products are frequently conflated with GIS tools, though the two serve distinct purposes: EO provides the measurements, while GIS provides the framework for managing and analyzing spatial data more broadly; see Box 8.1.). In an agro-meteorological context, this may involve estimating the volume of precipitation at a given location and time, inferring surface soil moisture conditions shortly after planting, or approximating vegetation growth just before harvest.

Yet, EO-derived datasets can differ along several dimensions. Chief among these differences is the type of signal observed by EO sensors, such as energy emitted by the land surface, radar signals scattered back from the surface, or sunlight reflected across different wavelengths (spectral bands). Sensors can also vary in the resolution of their observations, including how finely they can distinguish features on the ground (spatial resolution), how often they revisit the same location (temporal resolution), the wavelengths they measure (spectral resolution), and how subtle a difference in energy they can detect (radiometric resolution). Finally, EO-derived datasets can differ in how the raw observations are cleaned, adjusted, combined with other sources of information, and extended across locations or periods where observations are missing. These variations create many plausible datasets across temporal, spatial, and spectral dimensions, such that products intended to capture the same environmental phenomenon can diverge in practice and even appear to disagree on an underlying ``objective'' quantity.

For empirical research, these differences can be consequential. Seemingly innocuous choices about which EO product to use can affect both the magnitude and, at times, the sign of estimated regression coefficients. In this sense, selecting an EO dataset amounts to choosing one DGP from a set of relevant but imperfectly substitutable alternatives. As with other data choices made in empirical research, the selection of an EO product can shape subsequent analytical decisions, including how variables are constructed or aggregated and which identifying assumptions are required for causal inference.

\begin{chapbox} \textbf{Box 8.1. Distinguishing between GIS tools and EO data}

Although often used together, geographic information system (GIS) tools and Earth observation (EO) data serve distinct but complementary roles in geospatial analysis. EO refers to measurements of the Earth's surface and atmosphere collected via airborne and spaceborne sensors and subsequently processed into variables such as temperature, precipitation, and vegetation indices (e.g., the normalized difference vegetation index, NDVI). EO datasets are typically organized as grids of measurements covering an area (rasters), with sensors (and the products derived from them) differing in how much detail they capture, how often they revisit the same place, and what they detect. \\

GIS tools, in contrast, provide a framework for storing, managing, analyzing, integrating, and visualizing spatial data. They can incorporate both EO-derived inputs and non-EO spatial data, such as administrative boundaries, roads, or survey locations. GIS tasks are often carried out using desktop GIS platforms such as ArcGIS or QGIS, as well as cloud or programming-based tools such as Google Earth Engine or R. By integrating diverse datasets (e.g., raster or vector data derived from remote or ground-based sensors), GIS tools allow social scientists to, for example, link household surveys with environmental conditions and map geographic patterns in inputs and outcomes. Together, EO and GIS tools enable rigorous GIEs by connecting environmental measurements with socioeconomic data (see Campbell et al. 2022 for further discussion).
\end{chapbox}

\subsection{Types of gridded weather data}

Understanding the data generating process underlying weather datasets used in geospatial impact evaluations is crucial because weather products do not simply record measured ``weather.'' Instead, they operationalize specific constructs, such as rainfall measured at a gauge station, estimated precipitation over a grid cell, near-surface air temperature, or land surface temperature, through different combinations of observations, models, interpolation, and aggregation. The assumptions and processing decisions embedded in these products therefore shape how well a given product reflects the construct of interest.

One useful organizing framework distinguishes weather datasets according to how observations are turned into gridded measures, with key implications for measurement and inference in applied research (Auffhammer et al. 2013). Moving from products that are more directly observation-driven to those that more fully integrate statistical or physical modeling, weather datasets can be grouped into four categories: (1) \textbf{station-based products}, which rely on in situ observations and spatial interpolation; (2) \textbf{satellite retrievals}, which infer weather variables from remotely sensed data; (3) \textbf{merged products}, which combine multiple observational sources, often satellite estimates and ground observations; and (4) \textbf{reanalysis products}, which assimilate many types of observations into weather models to produce consistent and complete reconstructions of past conditions.

Across these categories, two recurring considerations tend to affect data quality and should be kept in mind throughout this section. First, products that use gauge observations can inherit limitations from the underlying station network: Errors often increase where stations are sparse, unevenly distributed, or poorly maintained. Second, satellite- and model-based products rely on retrieval, blending, or assimilation algorithms that introduce assumptions and potential biases, which can be especially consequential in complex terrain and in settings where weather patterns are often highly localized. For applied research, these construction choices generally create tradeoffs between smoothness and local detail, especially in how well products capture extremes and short-term variation.

\subsubsection{Station-based products}

Station-based products begin with ground observations, such as rain gauge measurements. However, because gauges measure precipitation only at specific locations, station-based products must rely on interpolation to estimate values between gauges and create continuous gridded surfaces. As a result, the value assigned to any given grid cell depends on both the underlying station observations and on the method used to translate those point measurements into a surface. Some interpolation approaches give greater weight to nearby stations (e.g., inverse-distance weighting), while others fit continuous surfaces that smooth values across space and avoid abrupt jumps between locations (e.g., thin-plate splines), or incorporate geographic predictors such as elevation to account for systematic variation in precipitation values (e.g., regression kriging or models that incorporate climate and elevation information, such as PRISM).

These products offer several important advantages: Gauge measurements are often treated as a gold standard for local weather observations, and many station records for variables such as precipitation extend back decades or even more than a century (Menne et al. 2012), enabling long-run analyses that predate the satellite era. However, gauges are not error-free: Changes in gauge type, observer practices, or surrounding environmental conditions can affect the long-run consistency of records. The reliability of gridded products generated from them also depends on the density and distribution of the underlying station network. Importantly, station locations are not random. Their placement reflects operational needs, accessibility, security conditions, maintenance constraints, and the intended use of observations. As a result, stations are often concentrated near airports, roads, population centers, or economically important areas (WMO 2025), while rural, remote, conflict-affected, or topographically complex regions remain more sparsely monitored.

These siting patterns have important implications for impact evaluation. Interpolation tends to smooth spatial variability, which can attenuate estimated effects when evaluating responses to localized shocks, especially in areas located far from stations. Measurement error may also be non-classical if station coverage is correlated with remoteness, infrastructure, or economic activity, causing weather estimates to vary systematically across contexts. Under such conditions, the assumption that weather measurements are exogenous to outcomes of interest may not hold (e.g., as illustrated by Schultz and Mankin 2019). In short, remote, conflict-affected, or topographically complex regions are often underrepresented in station networks, making station-based products less reliable in some of the very places where gridded weather data are most needed due to challenging on-the-ground collection conditions.

\subsubsection{Satellite-based products}

Satellite-derived weather products begin with remote observations of the atmosphere or land surface. Sensors capture data related to cloud properties, atmospheric moisture, precipitation, snow, vegetation, soil moisture, land surface temperature, or surface energy balance. Retrieval algorithms then convert these observations into estimates of weather-relevant variables using physical relationships, statistical models, or combinations of both.

A key advantage of satellite-based products is their broad and spatially consistent coverage, especially in regions with sparse or poorly maintained station networks. At the same time, many weather variables are estimated only indirectly from satellite observations. For precipitation, infrared sensors may infer rainfall from cloud-top temperature, while microwave sensors rely on information about water and ice particles in clouds (Stephens et al. 2007). For temperature, satellites often observe land surface temperature rather than the near-surface air temperature commonly used in other modeling applications. For soil moisture, sensors detect microwave emissions or backscatter from the upper soil layer, which may not reflect deeper root-zone conditions. Moreover, the relationship between what satellites observe and the weather variable of interest can vary across regions, seasons, land cover types, terrain, atmospheric conditions, and times of day.

For empirical analysis, the main implication is that satellite-derived measures may not align exactly with the weather conditions most relevant for outcomes such as crop yields, labor productivity, or health. They may also differ both from the underlying construct of interest and from alternative measures of that same intended construct, such as those derived from station-based or merged weather products that capture near-surface conditions (e.g., at \textasciitilde2 m height). This introduces measurement error, often of the systematic (non-classical) rather than classical type, that can affect coefficient magnitude and interpretation, particularly when analyses focus on localized or short-duration weather shocks.

\subsubsection{Merged products}

Merged weather products combine satellite-derived estimates with ground-based gauge observations to leverage the complementary strengths of each source. Satellites provide frequent, spatially continuous coverage, while gauges provide direct surface observations that are often used to calibrate or bias-correct gridded products. Well-known examples include Climate Hazards Group InfraRed Precipitation with Station data (CHIRPS), which blends infrared satellite estimates with station data to produce quasi-global rainfall time series (Funk et al. 2015), and the Integrated Multi-satellite Retrievals for the Global Precipitation Measurement (GPM) mission (IMERG), which combines microwave and infrared satellite retrievals with gauge-based adjustments (Huffman et al. 2020).

To integrate these data sources, merged products rely on statistical and geospatial methods that combine multiple datasets into a single gridded estimate. Common approaches include regression kriging (which combines regression models with interpolation), optimal interpolation (which weights observations based on distance and expected error), or climatological scaling (which aligns satellite estimates with local long-run averages derived from gauge observations).

As with other weather products, these methods introduce assumptions that shape the resulting estimates. Merging requires reconciling differences in (a) spatial and temporal resolution and (b) what each source measures (e.g., cloud properties versus surface accumulation for precipitation). The performance of merging algorithms can also differ across contexts, particularly in regions with complex terrain or highly localized precipitation processes (e.g., convective rainfall in monsoon or semi-arid systems), where both satellite retrievals and gauge interpolation may be less reliable. As a result, merged products may exhibit location-dependent accuracy that reflects a combination of input data quality and modeling assumptions.

For empirical applications, merged products are widely used, especially for precipitation, because they often reduce bias relative to satellite-only estimates while maintaining the broad coverage of satellites. This can be especially valuable in certain types of econometric designs, where effects are often identified from deviations in weather conditions within the same unit over time. In settings where station records are sparse, discontinuous, or unevenly distributed, merged products may provide more consistent measurement of this within-unit variation. At the same time, if merging quality depends on factors such as station density or terrain, the resulting measurement error may be systematic rather than random. This can lead to heterogeneous attenuation of estimated effects and complicate comparisons across regions or populations.

\subsubsection{Reanalysis(assimilation) products}

Reanalysis products combine observations from many sources (including satellites, ground stations, radiosondes, aircraft, and ships), with weather models to estimate past weather conditions. Unlike merged products, which primarily adjust and combine observed data, reanalysis uses a physics-based model of the atmosphere together with measurements from multiple sources to constrain the model over time (i.e., keep it aligned with real atmospheric conditions). This process produces complete, physically consistent datasets across variables such as temperature, precipitation, humidity, and wind. Prominent examples include the fifth-generation European Centre for Medium-Range Weather Forecasts (ECMWF) reanalysis (ERA5; Hersbach et al. 2020), National Aeronautics and Space Administration (NASA)'s Modern-Era Retrospective Analysis for Research and Applications version 2 (MERRA-2; Gelaro et al. 2017), and the National Centers for Environmental Prediction/National Center for Atmospheric Research Reanalysis (NCEP/NCAR Reanalysis; Kalnay et al. 1996).

These features have important implications for empirical applications. Reanalysis products are widely used when long, internally consistent time series are needed or when multiple weather variables must be jointly analyzed. However, because values are partially model-derived, they may smooth spatial and temporal variation, dampen extremes, or shift the timing of short-lived events (e.g., heat waves or intense rainfall). This can attenuate estimated effects in settings where identification relies on high-frequency or localized shocks. More generally, because reanalysis outputs reflect a combination of observations and model assumptions, measurement error may be structured rather than random, with implications for inference that differ from those associated with station-based or merged products.

While all weather datasets involve some form of interpolation or modeling, the source of that structure differs across product types. Station-based datasets primarily reflect the spatial interpolation of point measurements; satellite products rely on retrieval algorithms that translate radiance measurements into geophysical variables; and merged products combine multiple observed inputs with statistical adjustments. Reanalysis products, in contrast, impose structure through a dynamic atmospheric model, which can introduce coherence across variables and over space and time. As a result, measurement error in reanalysis products may be more systematically tied to model dynamics rather than solely to data gaps or retrieval errors, with distinct implications for inference.

The remainder of this section reviews commonly used sources for key weather variables, focusing on the strengths and limitations of various EO-derived products used to construct specific metrics. We provide guidance on selecting appropriate products based on the research question, geographic context, and unit of analysis. Table~\ref{tab:8-A1} in the Appendix summarizes the datasets discussed in this section, including original documentation and seminal resources, spatial and temporal resolution, strengths and cautions for use, and examples of economic studies that use each product. The practical takeaway is that weather products, even when nominally measuring the same phenomenon, can differ in consequential ways; the appropriate choice depends on what is being measured, where, how, and at what scale.

\begin{chapbox} \textbf{Box 8.2. Weather versus climate}

While often used interchangeably, weather and climate are distinct concepts. Weather refers to the state of the atmosphere in a given place over short periods, such as cumulative rainfall over a week, daily maximum temperature, or wind direction and speed during a storm. Climate, by contrast, describes the statistical properties of weather conditions in a given location over a longer period, often summarized using 30-year reference periods known as climate normals. Figure~\ref{fig:8-2} illustrates this distinction, showing year-to-year weather fluctuations against a slower underlying climate trend. \\

This distinction matters for impact evaluation because weather and climate measure different phenomena. Weather data are often used to measure short-term shocks or variation, such as a dry spell during germination, a heat wave during flowering, or an unusually wet harvest period. Climate data help characterize local baselines and longer-run risks, for example whether a location is typically hot, how variable rainfall tends to be, or whether extreme events are becoming more common. A one-season drought is a weather shock; a long-run shift toward hotter growing seasons or fewer but more intense rainfall events is a climate pattern. \\

The distinction also matters for research design. In many panel studies in this literature, researchers account for each location's typical conditions (e.g., by including location fixed effects) before estimating the relationship of interest. This means that stable differences across places, such as one district being generally hotter or drier than another, are removed from the comparison (i.e., absorbed by the location's fixed effects). The analysis therefore asks whether outcomes change when a given district is hotter or drier than usual, rather than whether hotter or drier districts differ from other districts. In these designs, the variation supporting causal interpretation (i.e., identification) comes mainly from within-location deviations from typical conditions. In other words, the estimate comes from comparing each location to its own normal. \\

At the same time, some economic studies use short-run weather variation to learn about climate impacts: Year-to-year or within-season deviations can provide plausibly exogenous shocks because they generate variation that is unrelated to local economic or social conditions and can therefore be linked more credibly to observed outcomes. Interpreting these estimates as climate impacts, however, requires additional assumptions, including that responses to short-run fluctuations approximate responses to longer-run changes, and that adaptation does not fundamentally alter those relationships over time (Dell et al, 2014).\\ 

In this chapter, we focus primarily on weather variation at two scales: within-season variation (intraseasonal variability), such as heat waves, dry spells, or rainfall timing within a growing season; and year-to-year variation (interannual variability), such as differences in drought severity across years. Climate nevertheless remains conceptually and practically important, especially as the baseline against which these shocks are defined and interpreted.

\begin{center}
  \includegraphics[width=\textwidth,height=0.75\textheight,keepaspectratio]{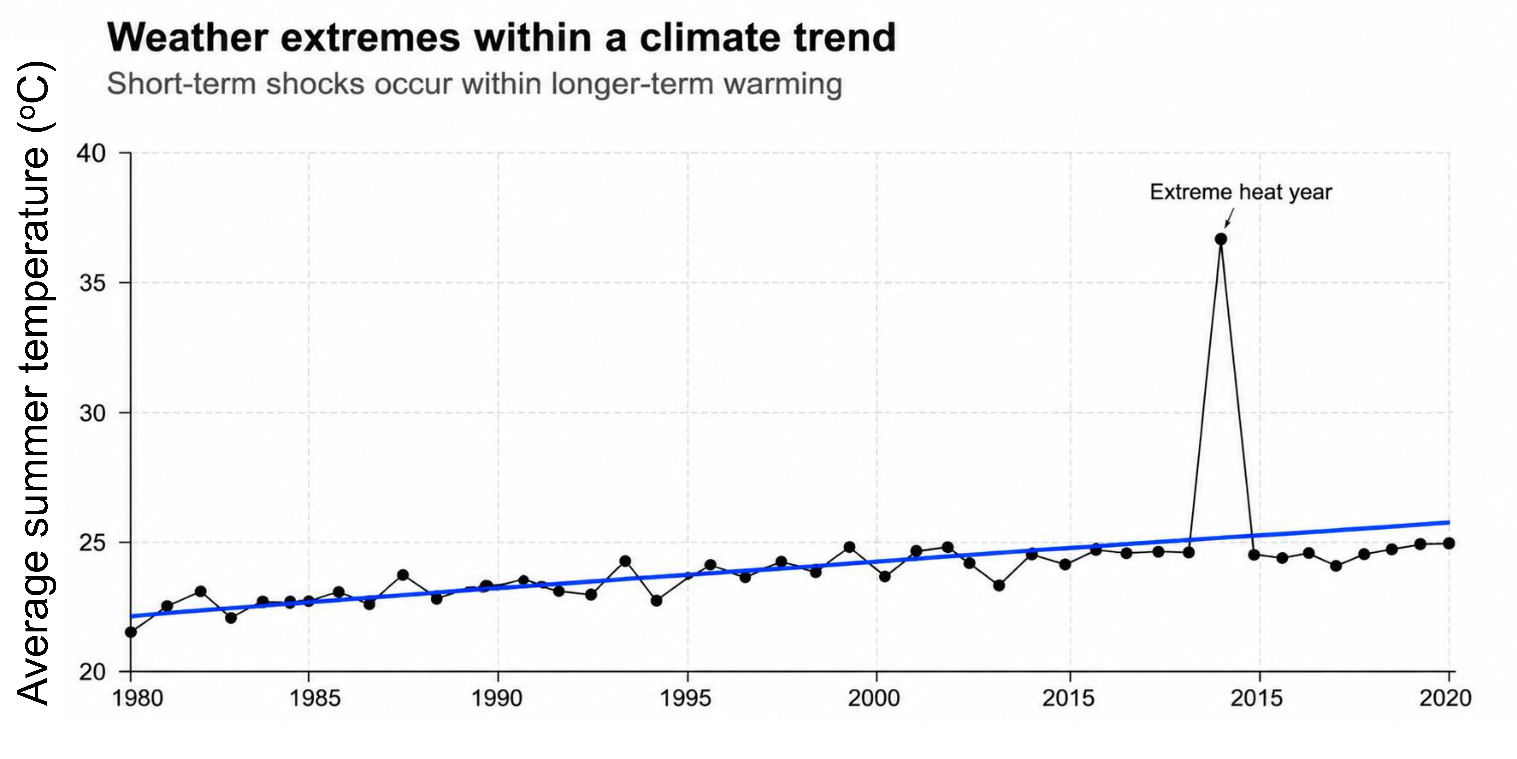}
  \captionof{figure}{Conceptual illustration of weather and climate. Year-to-year variation in average summer temperature (black points) fluctuates around a gradual upward climate trend (blue line). An extreme heat year, shown as a sharp departure from the trend, represents a rare but potentially high-impact deviation from typical conditions.}\label{fig:8-2}
\end{center}
\end{chapbox}

\subsection{Precipitation}

Precipitation is a common weather variable of interest for researchers conducting geospatial impact evaluations focused on economic development. This partly reflects the central role of agriculture in many low- and middle-income countries, where it remains an important source of livelihoods, food security, gross domestic product, employment, and household consumption. In addition, agricultural production often depends heavily on rainfall rather than irrigation. This is the case in low-income countries, where agriculture is almost exclusively rainfed (Rockström et al. 2007), but it is also true for much agricultural production in the Global North, where irrigation is often used only when rainfall is insufficient (Siebert et al. 2010). Notable exceptions include agricultural production west of the 100th in the United States (Seager et al 2018) or farming in arid regions such as the Negev Desert or western China. Compared with products measuring other weather phenomena, precipitation datasets are also often more readily accessible and easier to incorporate into geospatial analyses. For these reasons, rainfall is often one of the first variables researchers turn to when integrating Earth observation data into impact evaluations. Yet this combination of accessibility and ubiquity can breed complacency and reduce critical scrutiny of the different data generating processes underlying alternative products. Even for a commonly used metric like rainfall, the choice of data type and source can produce meaningly different results in an econometric model.

\subsubsection{Variation in EO precipitation products}

While precipitation encompasses all forms of atmospheric water deposition, including snowfall, hail, and sleet, rainfall (liquid precipitation) is the form most commonly measured and modeled in the empirical literature, and the two terms are often used interchangeably in practice. In some cases, this conflation is relatively harmless. For example, a researcher studying maize yields in lowland tropical areas may reasonably treat precipitation and rainfall as effectively equivalent, since the vast majority of precipitation in these environments falls as rain due to consistently warm temperatures.

The same assumption would be inappropriate, however, in settings such as the northern Great Plains of the United States or high-altitude agricultural regions (e.g., the Hindu Kush--Himalayan highlands or the Andes), where a substantial share of water inputs is stored as snow and released later through snowmelt. In these contexts, contemporaneous precipitation is a poor proxy for water availability, as soil moisture recharge, groundwater replenishment, and streamflow depend not only on the amount of precipitation, but also on the timing and magnitude of melt (Barnett et al. 2005; Immerzeel et al. 2010). Widely used rainfall-focused products like CHIRPS (Funk et al. 2015), which is specifically designed to capture liquid precipitation, may therefore systematically understate total water availability, potentially biasing estimates of outcomes for which total water supply is the relevant quantity. Researchers working in snow-affected environments should instead consider products that explicitly account for snow-water equivalent and snowmelt dynamics. For example, alternatives such as gridMET (Abatzoglou 2013) or the ERA5 reanalysis (Hersbach et al. 2020) are likely more appropriate than rainfall-focused products designed primarily for tropical and subtropical contexts.

Beyond the rainfall/precipitation distinction, credibly using rainfall for GIE means grappling with how different EO precipitation products construct their estimates, as seemingly small measurement and processing choices can translate into materially different results. Rainfall products vary along several key dimensions: their intended purpose (weather forecasting, climatological monitoring, etc.), how inputs are generated (station interpolation, satellite retrieval, reanalysis assimilation, or some combination thereof), and the spatial and temporal resolution at which outputs are made available. These differences often produce pronounced disagreement over what might otherwise appear to be an objective fact: how much rain fell in a given location over a given period. This problem is well documented in the hydrological and atmospheric science literature on precipitation product intercomparison (Beck et al. 2017; Sun et al. 2018), but it has received comparatively little attention in economics and the broader social sciences, where these products are increasingly applied.

\begin{figure}[htbp]
  \centering
  \includegraphics[width=\textwidth,height=0.75\textheight,keepaspectratio]{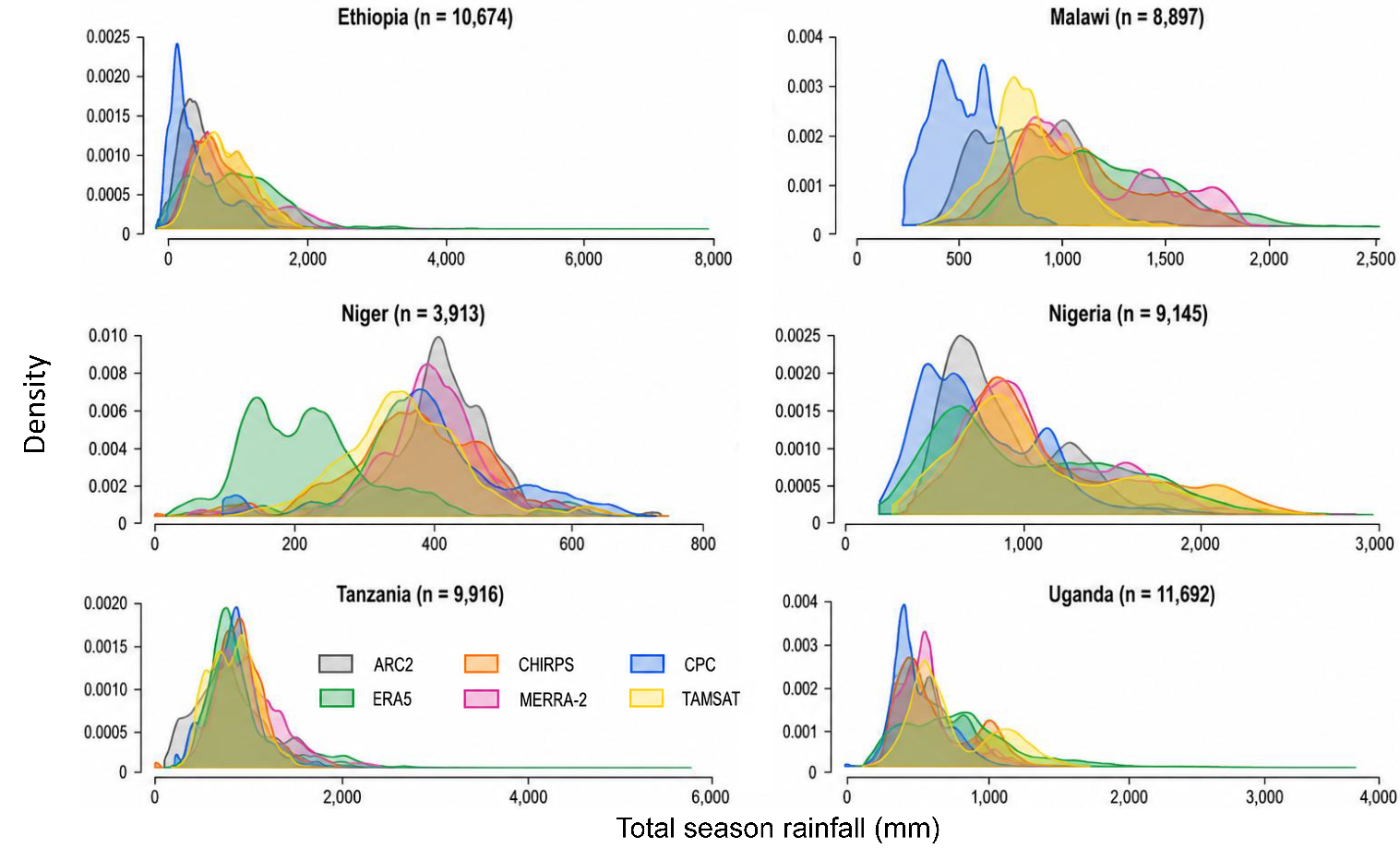}
  \caption{Total seasonal rainfall distributions for six countries in sub-Saharan Africa from six EO-based data products (Josephson et al. 2025). Distributions and tails differ substantially across products despite ostensibly measuring the same phenomenon.}
  \label{fig:8-3}
\end{figure}

Figure~\ref{fig:8-3}, from Josephson et al. (2025), illustrates how these differences in precipitation product types propagate across an entire growing season in six African countries. The pronounced disagreement among products has direct econometric consequences, since estimated results may reflect not only rainfall itself but also the particular rainfall data generating process selected by the researcher. This is precisely what Josephson et al. (2025) find: Researchers may obtain markedly different, or even oppositely signed, rainfall coefficients depending on the rainfall product used, a finding with sobering implications for parts of the existing literature. Every product type has its strengths and weaknesses, and researchers must therefore select among them based on careful consideration of what their research question requires of the data, rather than defaulting to whichever product is most familiar or readily accessible.

In addition to differences in how inputs are generated, EO precipitation products also differ in their temporal and spatial resolution. Products may report data at hourly, daily, dekadal (10-day), or monthly intervals, and grid cell sizes vary substantially, ranging from around 5 km for high-resolution satellite-based products commonly used in GIEs (e.g., CHIRPS) to 50 km or more for coarser reanalysis and gauge-based products (Sun et al. 2018)---though radar-based products available in data-rich countries can achieve even finer resolutions of approximately 1 km (e.g., the Next Generation Weather Radar, or NEXRAD, in the United States).

\begin{figure}[htbp]
  \centering
  \includegraphics[width=\textwidth,height=0.75\textheight,keepaspectratio]{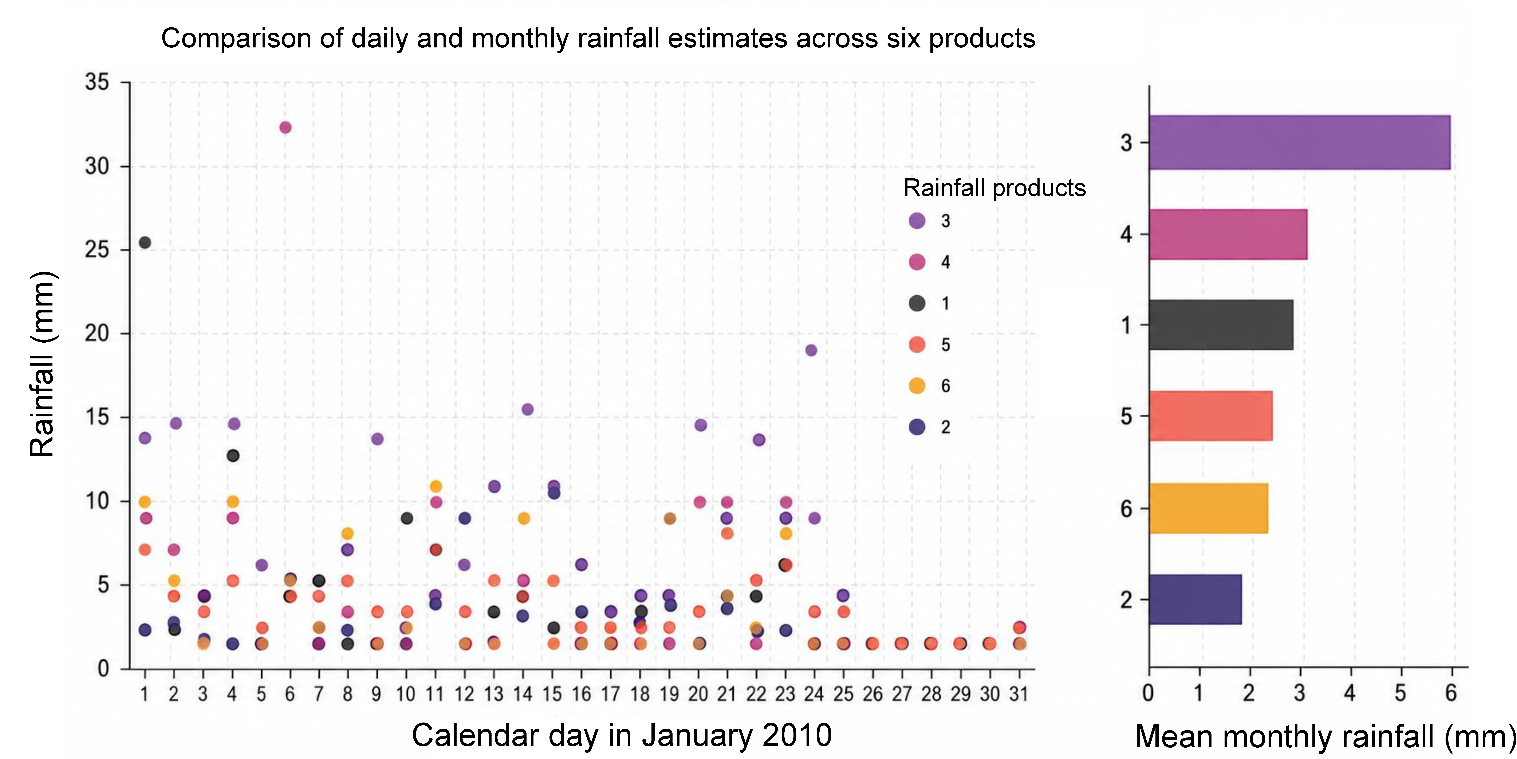}
  \caption{Daily rainfall estimates from six EO products (ARC2, CHIRPS, CPC, ERA5, MERRA-2, and TAMSAT) for the same location in January 2010 (left), with the mean daily rainfall for each product over the same month shown for comparison (right). On individual days, products can differ by an order of magnitude or more, and these differences also produce meaningfully different monthly averages.}
  \label{fig:8-4}
\end{figure}

Researchers might be inclined to assume that finer spatial and temporal resolution is inherently better, but this is not necessarily the case. The appropriate resolution depends on the research question: Hourly data may add little value if the outcome of interest responds to seasonal water availability, while daily or dekadal data may be insufficient for research questions involving extreme precipitation events or flash flooding, where within-day rainfall intensity is the relevant quantity (Kidd and Huffman 2011). Finer resolution also comes with statistical tradeoffs: As grid cells shrink and time steps shorten, variance in reported rainfall tends to increase, reflecting not only greater true meteorological variability, but also greater sensitivity to interpolation error and instrument noise. A product that appears precise at fine scales may therefore be less reliable than a coarser product that aggregates over more observations and smooths out these sources of error. Figure~\ref{fig:8-4} illustrates this point by comparing raw daily rainfall and mean daily rainfall from six EO products for the same day and location. Much of the day-to-day variation within any given product tends to smooth out over longer time horizons, highlighting that high temporal precision does not necessarily imply high accuracy at any given point in time.

The spatial structure of rainfall adds a further layer of complexity. Weather events are often highly spatially correlated, meaning that rainfall conditions at nearby locations tend to be similar. This enables interpolation algorithms to infer values between sparse observations. Consequently, some of the spatial detail visible in a high-resolution product may partly reflect interpolated estimates rather than true variation in underlying conditions (Chen et al. 2008). This is consequential because the degree of spatial interpolation required is itself a function of the product's DGP: Products that rely more heavily on interpolation between sparse station networks will tend to produce smoother spatial fields, regardless of the nominal resolution at which outputs are reported. The spatial structure of rainfall adds a further layer of complexity. Weather events are often highly spatially correlated, meaning that rainfall conditions at nearby locations tend to be similar. Interpolation algorithms exploit this correlation to estimate values between sparse observations, which is how gridded products achieve continuous coverage. However, this also means that the apparent spatial detail in a high-resolution product can be misleading: fine-scale features in the grid may reflect the smoothing assumptions of the interpolation rather than independently observed variation in underlying conditions

\begin{figure}[htbp]
  \centering
  \includegraphics[width=\textwidth,height=0.75\textheight,keepaspectratio]{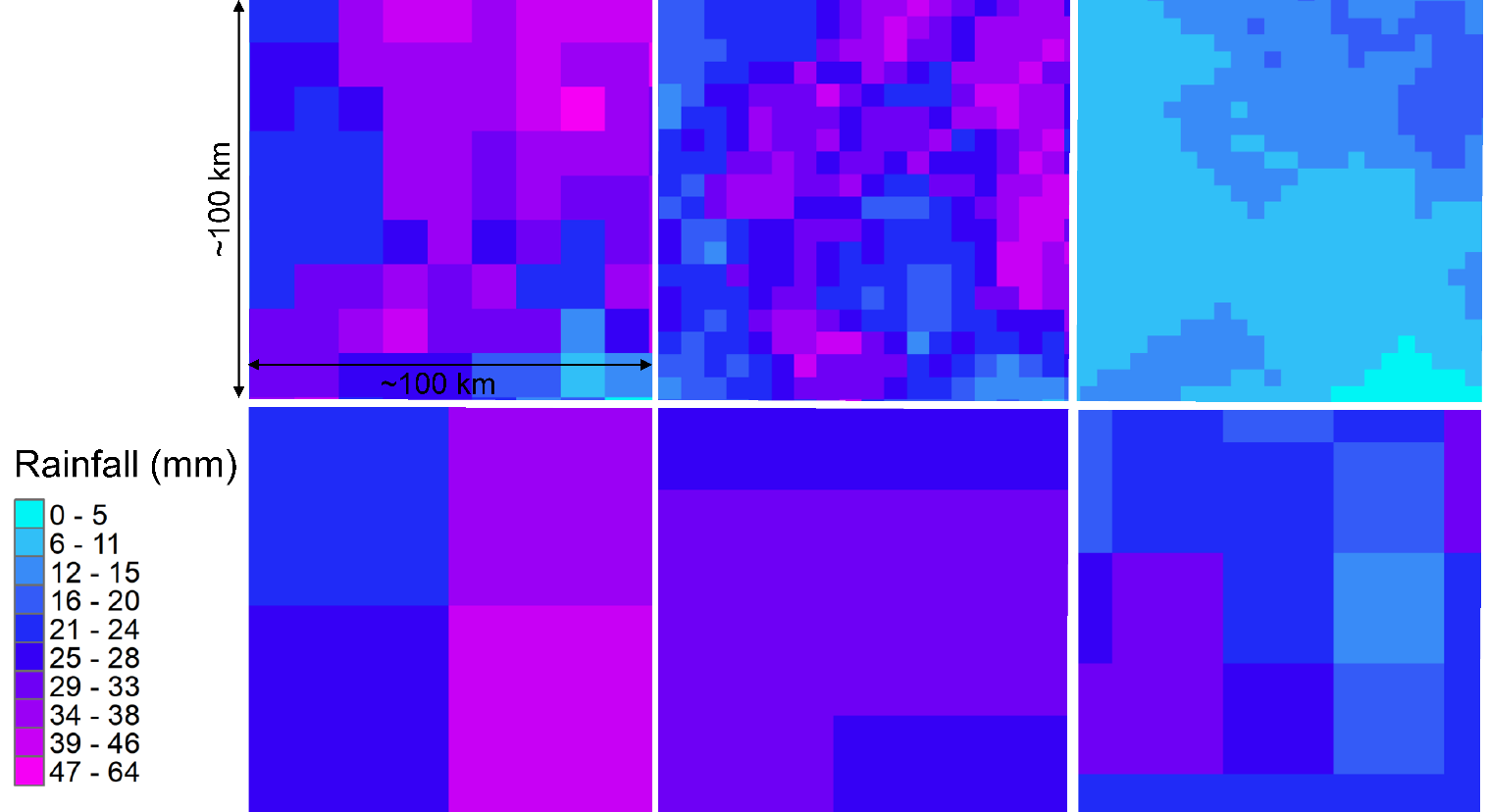}
  \caption{Varying spatial resolution of rainfall measurements across six EO products (from top left going clockwise: ARC2, CHIRPS, TAMSAT, CPC, MERRA-2, and ERA5) over the same 100 km × 100 km area on a single day (January 7, 2010) in Malawi. Image and code reproduced with permission from Siobhan Murray.}
  \label{fig:8-5}
\end{figure}

Figure~\ref{fig:8-4}, adapted from Michler et al. (2022), illustrates the practical stakes: For the same location on a single day, one product reports less than 5 mm of rainfall while another reports more than 47 mm. Such differences can lead to qualitatively different conclusions about exposure to a weather shock. Differences in reported rainfall and spatial resolution are often correlated precisely because resolution and interpolation method are not independent choices; both partly reflect the product's underlying DGP. Figure~\ref{fig:8-5} illustrates this visually by comparing the same 100 km × 100 km area in Malawi on a single day across six EO products. The varying grid cell size in each panel reflects the spatial resolution of the underlying product, while differences in the values reported within those cells illustrate how products diverge in their rainfall estimates for the same place and time.

\subsubsection{Precipitation product types}

A basic understanding of how each product type estimates rainfall informs which products are appropriate for a given research question.

\textbf{Interpolated station-based products} derive rainfall estimates from direct observations at weather stations and interpolate these across space to produce gridded outputs (Chen et al. 2008). These products are particularly useful for historical or long-term climate analyses, given that station records in some networks extend back a century or more. They can be highly accurate in regions with dense station networks, but these strengths often become limitations in low-income countries, where station coverage is sparse, unevenly distributed, and in some cases declining (Lorenz and Kunstmann 2012)---precisely the contexts in which GIEs are most commonly conducted. Widely used examples include the Climatic Research Unit Time Series (CRU) dataset developed by the University of East Anglia, the Global Precipitation Climatology Centre Monitoring Product (GPCC-MP), the National Oceanic and Atmospheric Administration Global Historical Climatology Network (NOAA GHCN), and the University of Delaware (UDel) Terrestrial Precipitation dataset, among others summarized in the Appendix.

\textbf{Satellite-based products} estimate rainfall indirectly, using sensor observations of clouds and atmospheric conditions to infer rain rate or accumulation via retrieval algorithms (Kidd and Huffman 2011). Because rainfall is not observed directly, these estimates are subject to retrieval uncertainty that varies by sensor type, algorithm, and meteorological context (Tian et al. 2009; Tian and Peters-Lidard 2010). Many satellite-based products were designed for operational monitoring and disaster response and are therefore available in near real-time at high temporal resolution (e.g., 30-minute intervals), making them better suited than coarser products for impact evaluations requiring measurement of short-duration or extreme rainfall events. However, retrieval uncertainty tends to be highest precisely during these events (AghaKouchak et al. 2012). Near real-time versions typically sacrifice some accuracy for timeliness, whereas climate data record (CDR) versions, which apply bias correction using station observations, are generally more appropriate for research applications (Becker et al. 2013). Examples include the Precipitation Estimation from Remotely Sensed Information using Artificial Neural Networks Climate Data Record (PERSIANN-CDR), the Climate Prediction Center (CPC) MORPHing technique Climate Data Record (CMORPH-CDR), and UC-Santa Barbara's Climate Hazards InfraRed Precipitation (CHIRP), among others summarized in the Appendix.

\textbf{Merged products} combine satellite-derived estimates with ground station observations, typically using station data to bias-correct or statistically blend satellite estimates and produce gridded outputs with broader spatial coverage than station networks alone could provide (Funk et al. 2015; Huffman et al. 2020). By combining satellite coverage and station-level precision, merged products often outperform purely satellite-based or purely interpolated products in regions with moderate station density. This advantage, however, tends to erode where station networks are sparse, which introduces variation in performance over space (Dinku et al. 2018). This performance heterogeneity constitutes a form of non-classical measurement error, in which errors are correlated with the true value of the variable being measured and vary systematically across contexts or conditions (Bound et al. 2001).

Merged products are among the most used precipitation datasets in GIEs. Widely used examples include the African Rainfall Climatology version 2 (ARC2), the Tropical Applications of Meteorology using SATellite data and ground-based observations (TAMSAT), the Integrated Multi-satellitE Retrievals for GPM (IMERG)---which succeeded the Tropical Rainfall Measuring Mission (TRMM) following the end of TRMM data collection in 2015---and the Climate Hazards InfraRed Precipitation with Stations (CHIRPS), among others summarized in the Appendix.

\textbf{Reanalysis products} differ fundamentally from the product types discussed above in that they are not primarily observational. Instead, they integrate multiple data streams, including satellite retrievals, ground station records, radiosondes, and increasingly radar and global positioning system (GPS) occultation data, into physics-based atmospheric models to produce internally consistent, globally complete estimates of precipitation and other meteorological variables (Dee et al. 2011; Gelaro et al. 2017). This model-based approach makes reanalysis products particularly well suited to research designs examining long-term climate variation and the joint effects of multiple weather variables, such as rainfall, temperature, humidity, and wind. However, because reanalyses are optimized for global atmospheric consistency rather than local accuracy, their spatial and temporal resolution may be too coarse to capture extreme events, microclimates, or farm- and household-level variability; that is, precisely the types of variation relevant to many impact evaluations (Parker 2016). These limitations notwithstanding, reanalysis products are uniquely valuable for research designs requiring globally consistent, long-run weather records or exploiting large-scale climate variation as a source of identifying variation; in such contexts, the spatial smoothing inherent in reanalysis may be an asset rather than a limitation.

Widely used examples include ERA5, MERRA-2, NOAA's Climate Forecast System Reanalysis (CFSR), and the Japanese 55-year Reanalysis (JRA-55), among others summarized in Table~\ref{tab:8-A1}.

In summary, precipitation occupies a central place in geospatial impact evaluations for good reason: It is a critical input for agricultural productivity in the predominantly rainfed farming systems where development-focused research is most commonly conducted, and rainfall datasets are among the most accessible EO products available. Yet accessibility can obscure complexity, and selecting a precipitation product is not a merely technical choice. Product selection can alter both the magnitude and direction of estimated effects, while mismatches between product characteristics and research context can introduce systematic errors with important consequences in high-stakes applications such as impact evaluation, early warning systems, or insurance design.

Researchers should therefore select precipitation products deliberately, matching spatial and temporal resolution, generation method, and geographic coverage to the demands of their research question. They should also consult the documentation accompanying each product before use, paying particular attention to period of record, known performance in the region of interest, and prior use in similar applications. In all contexts, careful review of each product's documentation is essential to ensure that data are fit for purpose before they are incorporated into empirical analysis.

\subsection{Temperature}

Temperature is among the most widely used weather variables in GIEs and is often equally or more important than precipitation for explaining variation in both agricultural and non-agricultural outcomes. In agriculture, temperature strongly influences productivity through its effects on crop development, water demand, and heat stress, particularly in temperate zones where seasonal and diurnal temperature fluctuations are large (Lobell and Field 2007; Lobell et al. 2011). Temperature also affects productivity through nonlinear threshold effects, a pattern observed across many climatic zones (Schlenker and Roberts 2009). Beyond agriculture, temperature influences labor productivity, learning, and other dimensions of human capital formation (Graff Zivin and Neidell 2014; Shah and Steinberg 2017). As global mean temperatures continue to rise, these effects are expected to intensify. Evidence suggests that the most severe impacts on agricultural productivity will be concentrated in Africa, Asia, and Central and South America (Deryng et al. 2014; Ortiz-Bobea et al. 2021), though recent research also documents substantial negative effects in the Global North (Carleton et al. 2016; Schauberger et al. 2017; Ortiz-Bobea et al. 2025).

\subsubsection{Variation in EO temperature products}

\begin{figure}[htbp]
  \centering
  \includegraphics[width=\textwidth,height=0.75\textheight,keepaspectratio]{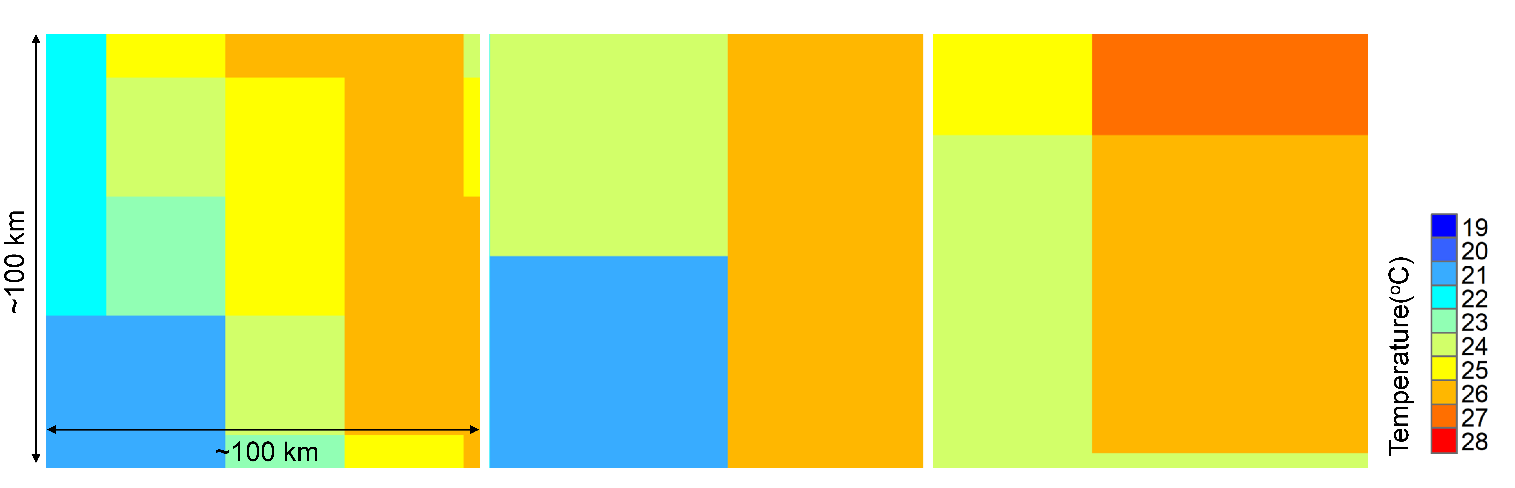}
  \caption{Varying spatial resolution of temperature measurements across three EO products (from left to right: CPC, MERRA-2, and ERA5) over the same 100 km × 100 km area on a single day (January 7, 2010) in Malawi. Image and code reproduced with permission from Siobhan Murray.}
  \label{fig:8-6}
\end{figure}

Given temperature's relevance across a wide range of outcomes, researchers must understand what an EO temperature measure represents, since products differ in both what they observe and how these observations are processed into gridded estimates. As with precipitation, remotely sensed temperature data can exhibit substantial heterogeneity across products for the same location and time period, as illustrated in Figure~\ref{fig:8-6} for three commonly used products in Malawi.

Because raw temperature is not always the most policy-relevant measure, researchers often construct agro-climatic indicators that better capture the mechanisms linking temperature to outcomes of interest. The most widely used are growing degree days (GDD), defined as accumulated heat exposure above a crop-specific base temperature required for crop development, and killing degree days (KDD), defined as accumulated exposure above a high-temperature threshold at which heat begins to damage crop growth or yields (Ritchie and Nesmith 1991; Schlenker and Roberts 2009).

Like precipitation datasets, temperature products fall into four broad categories based on how estimates are generated: interpolated station data, satellite-based data, merged data, and reanalysis data. Products differ in both spatial and temporal resolution, and these differences can be consequential for empirical analysis. Heat-related mortality, for instance, may respond to multi-day heatwaves, requiring daily maximum temperatures and measures of consecutive days above critical thresholds, whereas analyses of long-run adaptation or climate trends may need only monthly or seasonal averages.

Beyond resolution, temperature measures raise two additional issues that researchers should consider when selecting and applying EO products. The first concerns the timing of the observation. Unlike precipitation, which accumulates over time and can be summed across intervals, temperature is a state variable (i.e., a snapshot of atmospheric or surface conditions at a given moment), and daily means may obscure the exposure most relevant to the outcome of interest. In agricultural and biological applications, maximum temperature, minimum temperature, chilling hours, or separate daytime and nighttime temperatures may be more informative than a simple daily average, since biological responses to temperature are often nonlinear and tied to physiologically meaningful thresholds rather than mean conditions (Hatfield and Prueger 2015). Similarly, for labor productivity and health outcomes, peak daytime temperatures or overnight minimum temperatures may matter more than daily means (Graff and Neidell 2014).

The second issue concerns where in the air--surface profile the temperature is measured. Many weather station records and reanalysis products report near-surface air temperature, conventionally measured at 2 m above the ground in a shielded instrument shelter, following World Meteorological Organization guidance (WMO 2025). Satellite-based products, by contrast, typically retrieve land surface temperature (LST), which reflects the thermal condition of the land surface itself rather than the overlying air. These are related but distinct quantities: Land surface temperatures can differ from near-surface air temperatures by several degrees (and sometimes by 10°C or more), particularly during hot, dry daytime conditions when radiative heating of the surface is intense and evaporative cooling is limited (Mildrexler et al. 2011; Good et al. 2017).

Near-surface air temperature is most commonly used in agronomic and economic crop--weather models, in part because long-run station records are widely available and because most established response functions were originally estimated using near-surface air temperature. Land surface temperature, however, may better capture surface energy balance, soil moisture stress, urban heat exposure, vegetation stress, or wildfire-relevant conditions. It is therefore more appropriate when the hypothesized mechanism operates through the thermal condition of the land surface rather than atmospheric exposure. The choice between the two should be guided by the mechanism hypothesized to drive the outcome of interest.

\subsubsection{Temperature product types}

\textbf{Interpolated station-based products} report near-surface air temperature, conventionally measured at 2 m above the ground (T2m), along with daily and monthly minimum, maximum, and mean temperatures. As with their precipitation counterparts, their accuracy is a function of station network density, and spatial resolution varies widely, from several hundred meters to 50 km. Many organizations that produce gridded precipitation products also produce gridded temperature products using analogous methods. Widely used examples include the CPC Global Unified Temperature dataset, the CRU dataset, NOAA's GHCN, WorldClim, and Udel's Terrestrial Air Temperature dataset, among others summarized in the Appendix.

\textbf{Satellite-based temperature products} differ from interpolated station products in several important ways. First, they retrieve land surface temperature (LST) rather than near-surface air temperature (T2m). As discussed above, these are related but distinct quantities, and the relationship between them varies by location, land surface type, vegetation cover, and elevation, and tends to be stronger at night than during the day (Good et al. 2017; Benali et al. 2012). Researchers requiring near-surface air temperature can estimate it from satellite-derived LST, though doing so requires local station data for calibration (Benali et al. 2012; Good et al. 2017). A further practical limitation is that satellite LST retrievals are only available under cloud-free conditions, which can introduce systematic gaps in coverage, particularly in tropical regions with persistent cloud cover or during rainy seasons when temperature--crop interactions may be most consequential (Good et al. 2017).

Second, satellite-based products provide temperature readings at fixed overpass times rather than continuous observation, which limits their ability to capture true daily maxima and minima. NASA\textquotesingle s Moderate Resolution Imaging Spectroradiometer (MODIS) instrument, now largely superseded by NASA Visible Infrared Imaging Radiometer Suite (VIIRS), provided up to four retrievals per day at fixed overpass times (Terra at approximately 10:30 a.m./p.m. and Aqua at approximately 1:30 p.m./a.m. local solar time), thereby approximating but not directly measuring daytime maxima and nighttime minima. VIIRS, which flies on the Suomi NPP and NOAA-20 satellites, continues this approach at similar overpass times and comparable spatial resolution (Hulley et al. 2017). Aside from MODIS and VIIRS, widely used satellite-based temperature products include the United States Geological Survey (USGS) Landsat Surface Temperature collections and the European Space Agency's Sentinel-3 Sea and Land Surface Temperature Radiometer (SLSTR), among others summarized in the Appendix.

\textbf{Merged temperature products} combine satellite-derived LST with station observations to produce gridded estimates of near-surface air temperature, thereby addressing key limitations of both source types. By calibrating LST against station records, merged products provide T2 estimates at higher spatial resolution than interpolated station products alone, while using satellite data to fill gaps left by sparse station networks. Merged products also tend to offer longer and more homogeneous time series than purely satellite-based products, in some cases extending back several decades. Temporal resolution is typically daily or monthly, and spatial resolution ranges from approximately 1 km to 0.5° (\textasciitilde55 km). This combination makes merged products potentially suited to studying fine-scale temperature anomalies, such as urban heat islands or localized cold air pooling, without requiring researchers to perform their own LST-to-T2 conversion.

A widely used example is the Climate Hazards Center InfraRed Temperature with Stations maximum temperature product (CHIRTSmax; Funk et al. 2019), which provides monthly Tmax estimates at approximately 5 km resolution by blending satellite thermal infrared observations with station data. Because minimum temperatures are more difficult to retrieve from satellite thermal infrared observations, CHIRTSmax is limited to Tmax. A daily version, CHIRTS-daily, extends this to provide both Tmax and Tmin using ERA5 diurnal temperature range estimates to disaggregate monthly CHIRTSmax values, making it a hybrid of merged and reanalysis approaches (Verdin et al. 2020). Other widely used merged products include the Berkeley Earth Surface Temperature record (BEST; Rohde and Hausfather 2020), among others summarized in the Appendix.

\textbf{Reanalysis temperature products} share the advantages described in the precipitation section, including global coverage, long temporal records, and internally consistent multivariable datasets, but they also offer two additional benefits that are particularly relevant for temperature applications. First, reanalysis products typically provide much higher temporal resolution than other product types. Some reanalysis products report temperature at hourly intervals (e.g., ERA5 and MERRA-2), while others report at six-hourly intervals (e.g., CFSR and JRA-55), in all cases providing finer temporal resolution than interpolated station or merged products, which are typically daily or monthly. This is valuable for research questions where the timing of temperature exposure matters, such as studies of heat stress during outdoor labor or standardized examinations (Garg et al. 2020; Graff Zivin and Neidell 2014).

Second, because reanalysis products may include temperature alongside precipitation, humidity, wind, and other variables derived from a consistent modeling framework, they offer a coherent multivariable dataset for researchers studying joint weather exposures. This can help reduce methodological inconsistencies that can arise when temperature is drawn from one product and precipitation from another. The most widely used reanalysis products for temperature are the same as those used for precipitation: ERA5, NASA's MERRA-2, NOAA's CFSR, and JRA-55, among others summarized in the Appendix.

As with precipitation, selecting a temperature product requires careful attention to the characteristics of the product as well as its underlying DGP. Temperature applications introduce additional considerations specific to this variable: whether the product reports near-surface air temperature or land surface temperature, which reported metric is most relevant to the outcome of interest (mean, maximum, minimum, or a derived indicator such as GDD or KDD), and the timing of data collection relative to the biological or behavioral mechanism being studied. These choices should, again, be guided by the hypothesized mechanism linking temperature to the outcome of interest. Reviewing how a product has been used in similar regions or applications can help reveal known limitations, establish a methodological benchmark, and anticipate processing challenges---although such comparisons should inform rather than substitute for careful assessment of whether the product's characteristics are well matched to the research question at hand.

\subsection{Additional weather metrics}

Beyond rainfall and temperature, several other EO-derived weather variables are available to researchers conducting impact evaluations and may be more directly relevant than rainfall or temperature for certain research questions. These variables often operate through distinct pathways of impact: wind affects natural hazards, dispersal processes (such as smoke, pollution, and airborne pests), and evapotranspiration; relative humidity shapes aridity, water stress, and thermal comfort; and solar radiation and cloud cover capture elements of the surface energy balance relevant to crop productivity and thermal stress. Most of these variables are available through reanalysis products, which offer the advantage of multivariable consistency discussed in the preceding sections.

\subsubsection{Wind}

Wind is relevant to impact evaluations in several contexts, including natural disaster studies (particularly tropical cyclones and other high-wind events), dispersal processes (smoke, pollution, and airborne pests such as locusts), and agricultural water availability through its effects on evapotranspiration and plant water loss (Allen et al. 1998).

Three considerations are particularly important for product selection. First, spatial resolution: capturing sustained winds driving smoke or pest dispersal may require only coarse spatial resolution, while measuring damaging gusts at the farm or household level may require much finer spatial detail. Second, temporal resolution: the relevant time step depends on the mechanism of interest. Hourly or sub-hourly data may be needed for storm damage studies, while daily or monthly averages may suffice for evapotranspiration applications. Third, measurement height: Many products report wind speeds at 10 m above the surface, the standard meteorological reference height for surface observations, though different applications will require different heights. Evapotranspiration calculations following FAO-56 use wind speed at 2 m, typically converted from 10 m observations, whereas wind damage studies may require different wind metrics altogether.

Station-based wind observations are highly sensitive to local site conditions. Nearby terrain, vegetation, and structures can substantially affect anemometer readings, making simple spatial interpolation of gauge data unreliable even where station coverage is adequate (Jiménez et al. 2010). Reanalysis products, which integrate station wind observations alongside other data streams into physically constrained atmospheric models, are therefore the standard approach for gridded wind data. Widely used reanalysis wind products include ERA5, MERRA-2, CFSR/CFSv2, and JRA-3Q (the successor to JRA-55). All provide 10 m wind speed estimates, with ERA5 offering comparatively fine spatial (\textasciitilde0.25°) and hourly temporal resolution.

\subsubsection{Relative humidity}

Relative humidity (RH) influences agricultural and socioeconomic outcomes in several ways. High humidity reduces the body\textquotesingle s ability to cool through evaporative sweating, making conditions feel hotter and more physically stressful at any given air temperature (Steadman 1979). High humidity heat has also been linked to reduced labor productivity (Kjellstrom et al. 2009; Dunne et al. 2013).

Most EO RH products report values at 2 m above the surface, which is the standard reference height for human, animal, and crop exposure. Station-based gridded products, including the Climatic Research Unit dataset (CRU, extending back to 1901) and the Met Office Hadley Centre Integrated Surface Database of Surface Humidity (HadISDH, extending back to the 1970s), provide monthly RH with long historical coverage. Satellite-based products such as the Atmospheric InfraRed Sounder Level 3 (AIRS L3) offer twice-daily retrievals but are subject to data gaps under heavy cloud cover, as discussed in the context of LST above. Reanalysis products including ERA5, MERRA-2, CFSR/CFSv2, and JRA-3Q also provide 2 m RH, with ERA5 again offering hourly temporal resolution and \textasciitilde0.25° spatial resolution.

\subsubsection{Solar radiation, cloud cover, and sunshine duration}

Solar radiation, cloud cover, and sunshine duration are three interconnected measures of the surface energy balance that influence photosynthesis, crop growth, and thermal stress. Greater cloud cover, measured as the fraction of the sky obscured by clouds, tends to reduce sunshine duration and attenuate incoming solar radiation. While these three variables are connected with one another, they are not interchangeable: Cloud cover is a coarse indicator of light availability, sunshine duration captures how long direct solar irradiance reaches the surface (typically measured in hours), and solar radiation measures the actual energy flux reaching the surface (commonly expressed in watts per square meter, W/m²). Solar radiation and temperature together determine the radiative component of heat stress (i.e., the warming of surfaces and organisms through direct and diffuse irradiance), which complements the humidity-related mechanisms discussed in the preceding section. For photosynthesis and agricultural productivity, solar radiation is the most directly relevant of the three, though cloud cover and sunshine duration are commonly used as proxies where direct radiation measurements are unavailable.

Among these variables, cloud cover is the most readily accessible. It appears in products such as the CRU dataset and reanalysis products including ERA5 and MERRA-2. Cloud cover can also be retrieved from satellite-based products such as MODIS and Landsat, or accessed directly from the International Satellite Cloud Climatology Project High-resolution Series (ISCCP-H). Sunshine duration is less commonly reported as a stand-alone variable. ERA5 provides it as a daily accumulation in hours, partly because astronomical day length (the interval between sunrise and sunset for a given location and date) can be calculated directly from geographic coordinates and calendar date. However, astronomical day length captures photoperiod rather than actual sunshine duration, which also depends on cloud conditions.

Solar radiation is available from interpolated station-based products such as WorldClim, as well as reanalysis products including ERA5 and MERRA-2. For research questions specifically focused on surface energy balance, dedicated products such as NASA's Global Energy and Water Exchanges Surface Radiation Budget Release 4 Integrated Product (GEWEX SRB R4-IP) and the Clouds and the Earth's Radiant Energy System Energy Balanced and Filled dataset Edition 4.1 (CERES EBAF Ed 4.1) provide solar radiation, cloud cover, and additional energy-related variables. GEWEX SRB is a merged product drawing on multiple inputs, including MODIS-derived retrievals and reanalysis data, while CERES EBAF is derived primarily from satellite observations (Loeb et al. 2018).

These products are of particular value to researchers seeking to go beyond temperature and precipitation to construct more physically grounded exposure measures. Applications include modeling crop yield responses to radiation variability, reconstructing physiologically relevant heat stress indices for health and labor analyses, estimating cooling demand and solar energy supply, and assessing differential climate vulnerability across income groups.

\subsubsection{Synthesis}

Wind, relative humidity, and solar and cloud-related variables can complement rainfall and temperature by capturing additional pathways through which weather affects human and agricultural outcomes, including hazard intensity, water balance, thermal comfort, and surface energy constraints on productivity. Variable selection should follow the hypothesized mechanism of impact. For storm damage or dispersal processes, prioritize wind; for agricultural productivity and labor outcomes under heat stress, consider relative humidity and solar radiation alongside temperature; and for questions involving photosynthesis and surface energy balance, combine solar radiation with cloud cover or sunshine duration.

Researchers should also remain attentive to the measurement limitations discussed throughout this section. Satellite-based retrievals of relative humidity and cloud-related variables can degrade under heavy cloud cover, while interpolated station products reflect the density and distribution of the underlying station network and, as noted previously, are subject to interpolation biases in data-sparse regions. As with rainfall and temperature, product choice can have meaningful impacts on empirical estimates. Where feasible, triangulating across product types (for example, comparing reanalysis with satellite retrievals) can help identify whether results are sensitive to the choice of data source and thereby strengthen confidence in the findings.

\subsection{Composite atmospheric indices}

While the measures discussed thus far capture what humans typically perceive as weather, these variables are not always the most relevant to plant or human health. Many outcomes of interest in impact evaluations instead depend on integrated physiological and biophysical stressors, which are often better represented by composite meteorological indices. Although these indices are not directly measurable as ``weather'' variables themselves, they are derived from standard meteorological observations and are designed to more closely capture mechanisms related to plant growth, heat stress, drought risk, and other environmentally mediated processes.

Composite indices can be grouped into three broad themes to guide selection and interpretation. Water availability metrics, including evapotranspiration and soil moisture, summarize how much water is available to plants and how quickly it is being depleted. Heat stress metrics, including wet bulb temperature and vapor pressure deficit, capture combined thermal and moisture stress relevant to both crops and human labor. Atmospheric condition metrics, particularly particulate matter concentrations, reflect exposure to degraded air quality associated with burning and pollution events that can affect health, productivity, and the reliability of other remotely sensed datasets.

Most composite indices can either be derived from the reanalysis products discussed above or accessed directly from multipurpose meteorological data portals. NASA's Prediction Of Worldwide Energy Resources (POWER) project draws on MERRA-2 and satellite-derived radiation products to provide solar and meteorological data across three focus areas (renewable energy, sustainable buildings, and agroclimatology) at hourly resolution but relatively coarse spatial resolution (1° × 1°). The AgERA5 product, derived from the ECMWF ERA5 Reanalysis and specifically designed for agricultural applications (Boogaard et al. 2020), provides daily agronomic variables at finer spatial resolution (0.1° × 0.1°). A revised version (v2.0) was released in May 2025, though as of January 2026 the dataset is provided on a best-effort basis only, with no guaranteed continuity of updates or investigation of data quality issues pending resolution of contractual arrangements with the external data providers (ECMWF 2025).

\subsubsection{Evapotranspiration}

Evapotranspiration (ET) measures the combined loss of water through evaporation from soil and open water surfaces and transpiration through plant leaves, making it a widely used proxy for plant-available water and a key input to drought and crop production analyses (Allen et al. 1998). Whereas variables such as precipitation primarily capture the supply side of the water balance, ET captures the demand side by reflecting how much water is being lost to the atmosphere and therefore how much needs to be replenished to sustain plant health and growth (Fisher et al. 2017). Because ET integrates precipitation, temperature, solar radiation, relative humidity, and wind, it serves as a natural composite exposure measure for impact evaluations in which outcomes depend on water balance rather than precipitation alone.

Water deficits that limit evapotranspiration below the level required for healthy crop growth are closely linked to reductions in crop yields, especially in rainfed agricultural systems (Doorenbos and Kassam 1979; Steduto et al. 2012). However, the strength of this relationship depends on both the timing and severity of the deficit within the growing season. ET is often most informative during early and middle stages of annual crop development, including leaf-area expansion, flowering, and grain filling, when water stress is especially likely to translate into yield losses.

Reanalysis-based ET products typically provide hourly or daily temporal coverage at spatial resolutions of several kilometers (Fisher et al. 2017). ERA5 and ERA5-Land provide actual ET and potential evapotranspiration (PET) derived from the model's land surface scheme at 0.25° and 0.1° resolution, respectively, while MERRA-2 provides ET at approximately 0.5° resolution (Hersbach et al. 2020; Gelaro et al. 2017).

Satellite-based ET products, including the Simplified Surface Energy Balance ((SSEBop) model; Senay et al. 2020) at approximately 1 km spatial resolution and the WaPOR database (Water Productivity through Open access of Remotely sensed derived data; FAO 2020; Mannaerts et al. 2020) at resolutions ranging from 30--250 m depending on coverage level, offer finer spatial detail than reanalysis products, though frequently at coarser temporal resolution, with ET variables available at dekadal, monthly, and annual intervals. For farm- or field-level applications, OpenET provides satellite-based ET estimates at 30 m resolution with daily cadence by ensembling (aggregating across) multiple ET models (Melton et al. 2022). Currently covering 48 US states, with expansion to Brazil underway, its relatively short period of record limits its applicability for long-run historical analyses. However, it represents a promising direction for ET measurement at scales relevant to individual farms and fields.

Overall, the two product types have complementary strengths: Reanalysis-based ET products provide temporal completeness and consistency over long time periods, while satellite-derived ET products offer finer spatial resolution that may better capture heterogeneity at the farm or landscape level. Researchers should choose between these product types based on whether temporal coverage or spatial resolution constitutes the binding constraint (i.e., the primary limiting factor) for the application at hand. For example, in a study of irrigation impacts across neighboring farms, spatial resolution may be the binding constraint because coarse pixels can obscure field-level differences. Researchers should also match the temporal window of ET exposure to the relevant stage of crop development.

\subsubsection{Soil moisture}

Soil moisture complements ET by capturing stored water rather than water loss, thereby providing an estimate of the water available to plants at a given point in time. Although soil moisture is a physical state variable, it reflects the cumulative influence of precipitation, temperature, evapotranspiration, and soil properties including texture, porosity, and organic matter content. In impact evaluations, soil moisture is commonly used as a proxy for drought conditions and water stress shocks (e.g., Sheffield and Wood 2008).

A key distinction in soil moisture measurement is the difference between near-surface soil moisture (roughly 0--5 cm depth) and root-zone soil moisture (typically 10--100 cm, depending on crop type, though this depth range may approach 200 cm for deeper-rooting crops), where crops draw most of their water. Near-surface moisture responds quickly to recent rainfall and is therefore especially informative for short-run droughts, while root-zone moisture more directly determines whether crops can sustain growth through dry spells (Allen 1998).

Many EO soil moisture products are designed for operational applications such as agricultural monitoring and food security forecasting (Bolten et al. 2009; Mladenova et al. 2020), offering relatively high temporal frequency at moderate spatial resolution. However, satellite-based passive microwave instruments are physically limited to sensing approximately the upper 5 cm of soil, therefore capturing near-surface conditions rather than root-zone moisture (10--100 cm). This distinction is important because root-zone moisture more directly governs crop water uptake and agricultural productivity.

Some satellite-based products address the surface-to-root-zone gap through data assimilation, using near-surface retrievals together with land surface models to estimate conditions at deeper soil layers, although accuracy varies across soil depths and environmental conditions (Babaeian et al. 2019; (Colliander et al. 2017). Researchers should therefore exercise caution when using satellite products as proxies for agricultural water stress (Dorigo et al. 2017); Entekhabi et al. 2010). Widely used examples include the NASA Soil Moisture Active Passive (SMAP) mission Level-4 9 km Soil Moisture product (SMAP L4-SM; Entekhabi et al. 2010) and the Advanced SCATterometer (ASCAT) 12.5 km soil moisture product (Wagner et al. 2013).

Reanalysis products provide root-zone moisture estimates that are often more relevant for plant growth, though typically at coarser spatial resolution and with less frequent updates than satellite-derived products. Depending on the research question, combining a satellite-derived soil moisture product for finer spatial detail with a reanalysis-based root-zone series for temporal completeness can provide a comprehensive picture of plant-available water.

\subsubsection{Wet bulb temperature}

While temperature and humidity have thus far been discussed separately, their joint effects on plant and human physiology are often more consequential than either variable alone, a relationship that wet bulb temperature (WBT) is specifically designed to capture (Tuholske et al., 2021; Wilson et al., 2024). Drawing its name from the wet bulb thermometer traditionally used to measure it, WBT reflects the lowest temperature achievable through evaporative cooling under current atmospheric conditions. For humans, WBT maps directly onto physiological heat limits: As humidity rises, the body's capacity to shed heat through perspiration declines, and extreme WBT values are associated with dangerous and potentially unsurvivable conditions (Kjellstrom et al. 2009; Raymond et al. 2020). For plants, WBT can serve as a useful indicator of combined thermal and moisture stress affecting transpiration dynamics.

Wet bulb temperature is not typically reported directly in most EO-derived or reanalysis datasets. Instead, it is usually derived from standard near-surface meteorological variables, especially air temperature and humidity, using psychrometric equations or approximations such as those proposed by Stull (2011). Reanalysis products like ERA5 and MERRA-2 provide the temperature and moisture fields needed to calculate WBT consistently across space and time.

\subsubsection{Vapor pressure deficit}

Vapor pressure deficit (VPD) measures the drying power of the air by quantifying the difference between the moisture the air could hold at saturation and the moisture it actually contains, thereby capturing both the thermal and hydric dimensions of atmospheric stress in a single variable. Higher VPD values indicate drier atmospheric conditions, which can simultaneously reduce water availability for plants and increase heat stress (Yuan et al. 2019). As such, VPD is often more mechanistically informative than temperature alone in settings where humidity varies substantially. VPD has been shown to be closely related to variability in burned forest areas (Seager et al. 2018) and is an established driver of plant physiological stress, including reduced photosynthesis, diminished growth, and increased mortality risk under high VPD conditions (Grossiord et al. 2020). Similar logic applies to human labor outcomes, where VPD may be a more physiologically meaningful measure of thermal stress than temperature alone because it captures the drying and cooling constraints that shape physical work capacity (Kjellstrom et al. 2009).

VPD responses can be nonlinear and crop-specific: The same change in VPD may have different implications across crop types and climatic zones, which affects how VDP effects should be modeled in impact estimations. VPD can be derived from reanalysis products or accessed directly through products such as the Gridded Surface Meteorological dataset (gridMET; Abatzoglou 2013), which also provides rainfall, temperature, wind, relative humidity, and ET at 4 km resolution with daily updates for the contiguous United States. For global coverage, TerraClimate (Abatzoglou et al. 2018) provides vapor pressure together with temperature, precipitation, wind, solar radiation, and climatic water balance at approximately 4 km monthly resolution from 1958 onward.

\subsubsection{Particulate matter}

Particulate matter (PM) is not a meteorological variable in the strict sense, but it is closely linked to weather-driven events (e.g., agricultural burning, deforestation clearing, wildfires, and industrial pollution) that generate spatially heterogeneous exposure shocks with documented effects on health, labor productivity, human capital formation, and agricultural production (Deryugina et al. 2019; Wen et al. 2022; Borgschulte et al. 2024). PM is typically measured as PM2.5 (particles $\leq$ 2.5 microns in diameter) or PM10 ($\leq$ 10 microns), with PM2.5 more commonly used in health and productivity studies because of its deeper penetration into the respiratory system (Pope and Dockery 2006) and its more consistent availability across monitoring and satellite-based datasets (e.g., van Donkelaar et al. (2021).

PM is relevant to EO-based impact evaluation in two important ways. First, it supports exposure mapping for health and productivity analyses. Second, PM contributes to atmospheric aerosol loading---that is, the concentration of small particles suspended in the atmosphere due either to natural processes (e.g., dust storms) or human activities (e.g., agricultural burning or industrial emissions). Heavy aerosol loads can confound spectral remote sensing in ways that should be considered when using satellite-derived outcome measures. More specifically, aerosols attenuate and scatter incoming solar radiation, affecting surface reflectance measurements across multiple spectral bands and introducing apparent changes in vegetation indices such as NDVI and EVI that reflect atmospheric conditions rather than true vegetation responses (Kaufman and Tanré 1992; Vermote et al. 1997). This issue is particularly consequential during burning seasons, when aerosol concentrations are highest and agricultural monitoring is ongoing, creating a potential confound between PM exposure and EO-derived outcome measures that researchers should account for when designing their analyses.

The most widely used satellite-derived PM2.5 dataset in the economics literature is the one developed by van Donkelaar et al. (2021), which provides global estimates at 0.01° (\textasciitilde1 km) resolution by combining aerosol optical depth retrievals from MODIS and MISR (the Multi-angle Imaging SpectroRadiometer) with chemical transport model simulations.

\subsubsection{Synthesis}

Composite atmospheric indices can add explanatory power in impact evaluations by aligning exposure measures more closely with the mechanisms through which weather affects outcomes. Water availability metrics such as ET and soil moisture are particularly informative for agricultural productivity and drought analyses, especially when exposure windows are matched to crop phenology. Heat stress metrics such as WBT and VPD are often more interpretable than temperature alone in humid or water-limited environments. PM can capture exposure to burning- or pollution-related shocks while also informing data quality considerations for EO-derived outcomes. As with the primary weather variables discussed earlier, product choice should follow the hypothesized mechanism linking weather to outcomes. Where feasible, triangulating across complementary products can strengthen confidence in results and help identify sensitivity to data source choice.

\section{EO data on vegetation}

A common pitfall in EO-based GIE is interpreting observed spatial or temporal differences in imagery as treatment effects when they may instead reflect variable construction choices. Consider land surface phenology (LSP), which often uses time-series analysis of vegetation indices to estimate key events in vegetation growth cycles. Even between MODIS and VIIRS, its purpose-built successor, LSP metrics for start- and end-of-season timing can exhibit persistent discrepancies, with biases of up to 10 days in agricultural landscapes (Moon et al. 2019). Since these metrics are widely used to model management events like planting dates, harvest timing, and crop maturity (Lobell et al. 2013; Diao 2020), sensor and index choice can introduce confounds that masquerade as treatment effects. The sections below introduce the conceptual and processing pipeline underlying EO-derived vegetation variables, as well as the stages within that pipeline where uncertainty may arise.

\subsection{EO data generation: Deriving variables from multispectral imagery}

\begin{figure}[htbp]
  \centering
  \includegraphics[width=\textwidth,height=0.75\textheight,keepaspectratio]{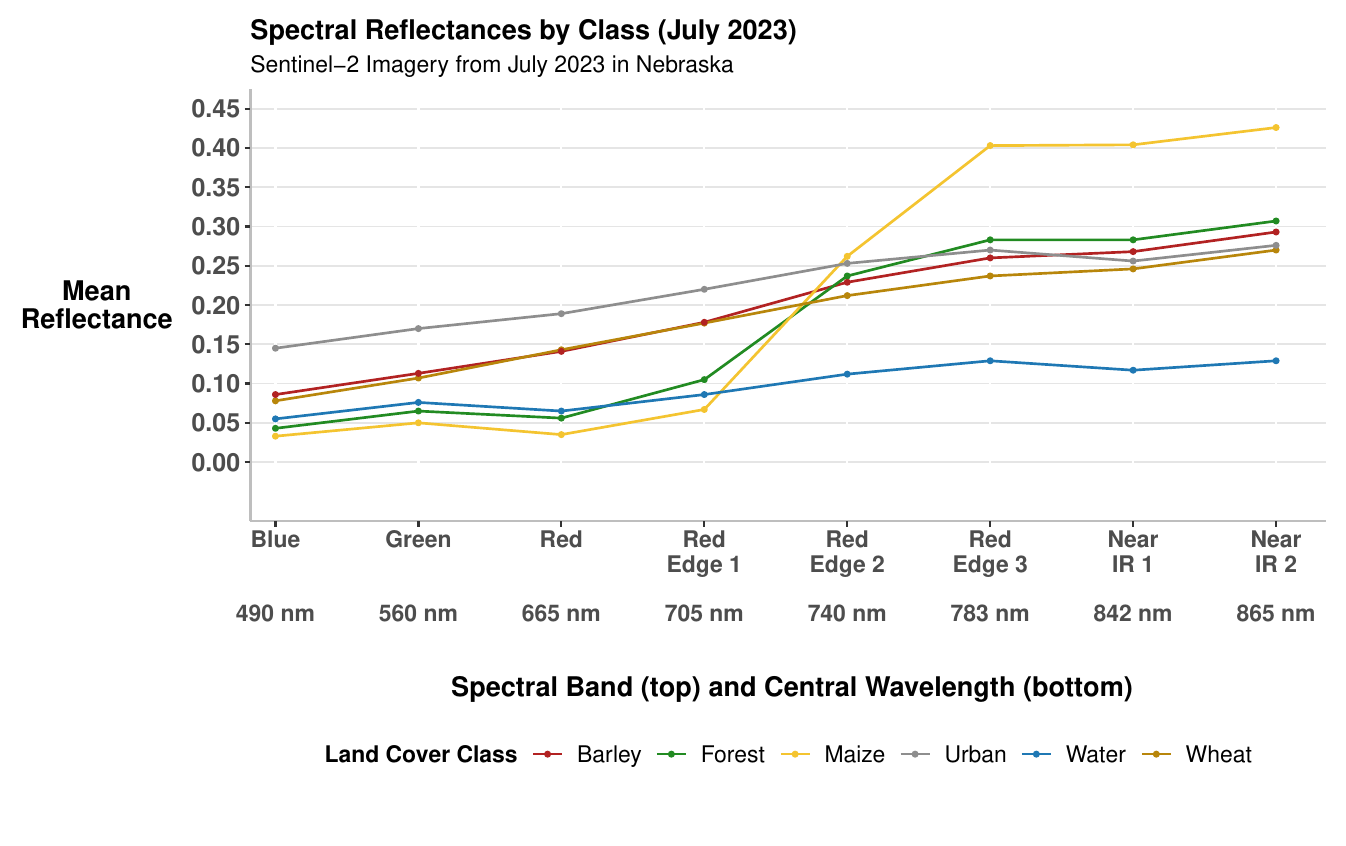}
  \caption{Spectral signatures from composite Sentinel-2 imagery for six sample Cropland Data Layer (CDL) land cover classes in Nebraska in July 2023, based on 50 sample points per class. Near IR = near infrared.}
  \label{fig:8-7}
\end{figure}

Satellites measure how different surface materials reflect or emit energy across discrete portions of the electromagnetic spectrum (EMS), known as spectral bands. Healthy vegetation, for example, strongly absorbs red light while reflecting strongly in the near-infrared (NIR) region, a contrast driven by chlorophyll absorption and internal leaf structure. This pattern is visible in Figure~\ref{fig:8-7}, where corn and wheat show markedly higher NIR reflectance than water while their visible-band values remain similar.

Stressed vegetation, bare soil, and water each exhibit different spectral responses, making it possible to infer surface condition through comparisons across spectral bands. The visible-to-near-infrared (VNIR) region has the longest history of application in agricultural remote sensing (Weiss et al., 2020), and forms the basis for the vegetation indices and land cover products discussed below.

Raw spectral measurements are generally used in three main ways: to compute spectral indices using mathematical formulas grounded in known biophysical relationships; to train classification models that assign pixels to land cover or crop type categories; and to conduct time-series analyses that track change over time. Many processed products derived from these approaches are freely available through space agencies and data providers, meaning that researchers do not always need to build products from scratch. Nonetheless, understanding the construction pipeline remains essential for interpreting what a given product does and does not measure.

\subsubsection{Resolution and classification accuracy.}

Spatial resolution shapes, though does not fully determine, classification quality. Resolution should broadly match the target object: Detecting individual tree crowns requires finer resolution than mapping broad forest types, for which coarser aggregated reflectance values are often more stable and less influenced by individual branches or leaves. Counterintuitively, resolution that is excessively fine relative to the target class can reduce accuracy by increasing within-class variability (Sadowski and Sarno 1976). Beyond spatial resolution, object properties such as contrast with the surrounding background, shape regularity, and spatial patterns also affect detectability. Narrow linear features like roads or irrigation ditches can remain detectable even when smaller than the nominal pixel size, since their high contrast and distinctive shape.

For change detection specifically, temporal consistency becomes at least as important as spatial resolution. Reliable detection of land cover change depends on frequent image acquisition, stable band calibration, and consistent geometric registration across dates (Townshend and Justice 1992). Differences in solar angle, atmospheric conditions, or sensor calibration between image acquisitions can generate apparent changes that reflect data artifacts rather than real surface dynamics.

\subsubsection{Vegetation indices.}

To reduce the complexity of multiband imagery, researchers calculate vegetation indices (VIs); that is, mathematical combinations of spectral bands designed to distill specific contrasts into a single interpretable value. The most widely used is the normalized difference vegetation index (NDVI; Rouse et al. 1974):

\begin{equation*} \mathrm{NDVI} = \frac{\mathrm{NIR} - \mathrm{Red}}{\mathrm{NIR} + \mathrm{Red}}
\end{equation*}

High NDVI values (above 0.5) generally indicate vigorous, photosynthetically active vegetation; moderate values (0.2--0.5) are more consistent with sparse vegetation, grasslands, or early-season crops; and values near zero or below typically suggest bare soil, built surfaces, or water. Importantly, NDVI does not include the green band despite its common description as a ``greenness'' index. Rather than measuring visible green color itself, NDVI captures photosynthetic activity through the contrast between red-light absorption and near-infrared reflectance. Values above 0.5 are commonly associated with healthy, dense vegetation, though these thresholds shift in smallholder systems, mixed-cropping environments, or areas imaged at coarser resolution, where mixed pixels blend multiple land cover types within a single observation.

\begin{figure}[htbp]
  \centering
  \includegraphics[width=\textwidth,height=0.75\textheight,keepaspectratio]{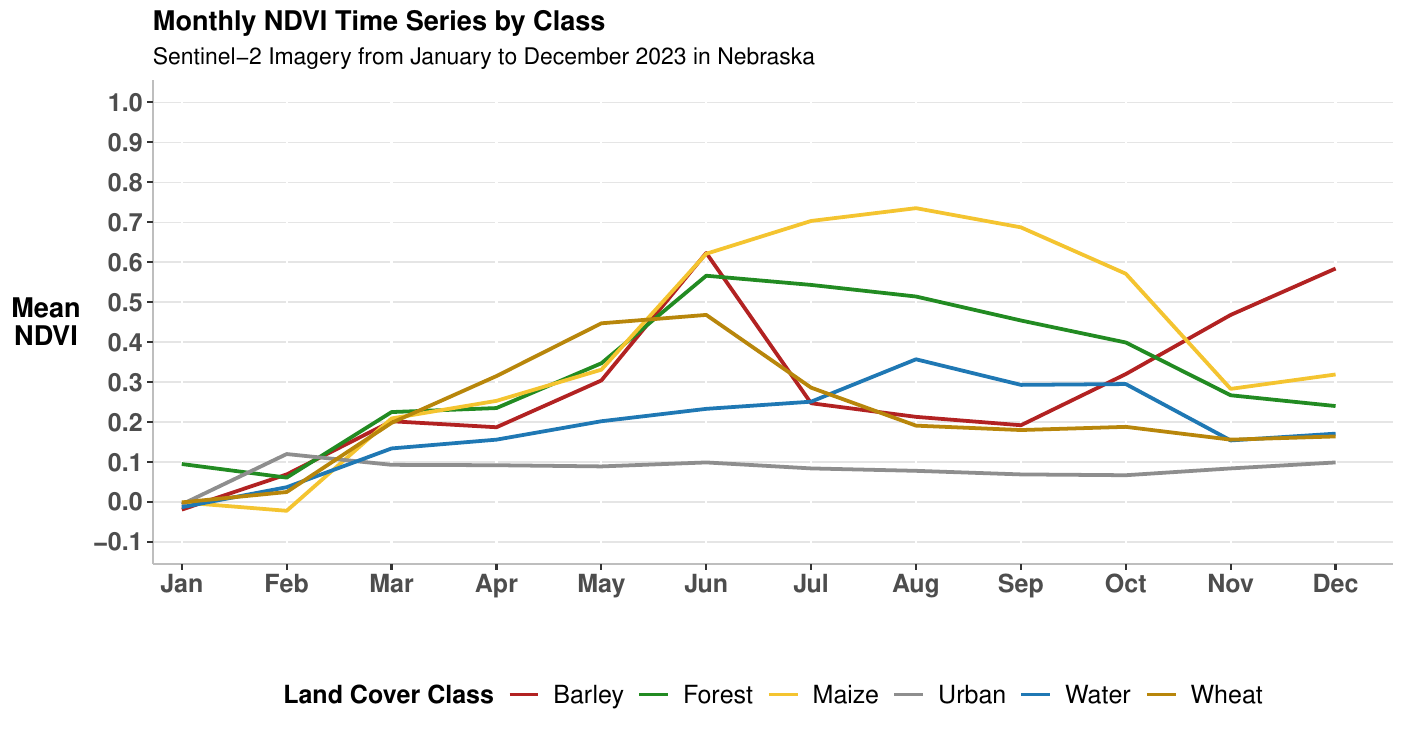}
  \caption{Monthly Sentinel-2 NDVI time series for a sample of observations drawn from six different CDL classes (2023). The time series are based on 50 sample points for each class. The separability of different classes depends on the month of year, field sizes, degree of intercropping, and issues such as cloud cover. Additional indices (beyond NDVI) can also improve class separability.}
  \label{fig:8-8}
\end{figure}

Figure~\ref{fig:8-8} illustrates how NDVI varies over the course of a calendar year across different land cover classes in Nebraska, with crops and forest reaching high values during the growing season while urban and water surfaces remain consistently low.

As a basic validation step, researchers should inspect a sample of locations to verify that observed NDVI patterns align with expected crop signatures and phenological dynamics. Some land type classification products, including the United States Department of Agriculture (USDA) Cropland Data Layer and WorldCereal, provide confidence bands indicating how well individual pixels conform to the model's classification rules as a gauge of classification performance (Boryan et al. 2011). These bands can help establish thresholds for change detection (Lark et al. 2021) or guide training data filtering in machine learning applications (Maleki et al. 2024).

\subsection{Selecting and interpreting vegetation indices and EO-derived variables}\label{selecting-and-interpreting-vegetation-indices-and-eo-derived-variables}

One of the most consequential decisions in integrating multispectral data into a GIE is selecting the appropriate product and index for the research question. The research context, including crop type, cropping system, field size, and agroclimatic zone in agricultural applications, helps determine which product and index are fit for purpose. The choice of index also depends on how the variable will be used. In predictive or classification tasks, the priority is maximizing class separability or predictive accuracy.

NDVI is perhaps the most used index in agricultural remote sensing. As Figure~\ref{fig:8-8} shows, it achieves meaningful separation between classes like corn and wheat during peak season in our sample case. However, that separation collapses during winter months, when all classes return similarly low values. NDVI-based classification is therefore highly sensitive to the selected time window and the underlying separability of the crops (low yielding wheat and barley crops may be more difficult to distinguish, for example).

In causal regression models, the priority shifts: NDVI must measure the specific biophysical quantity that the treatment is expected to affect. As a yield proxy, peak or integrated NDVI performs reasonably well in dense, uniform canopies in temperate environments, but saturates above moderate biomass levels and may fail to detect yield differences in some environments. The mapping between NDVI and yield varies across crop types. The index captures photosynthetic activity, but the relationship between NDVI and yield is shaped by multiple factors, including canopy architecture, growth habit, and phenological timing. As a result, a model calibrated for one crop may produce systematically biased estimates when applied to another.

This becomes particularly important when crop switching occurs. If treated and control plots differ in crop type as well as treatment status, differences in NDVI may reflect crop composition rather than the intervention itself, making it difficult to isolate a clean treatment signal. More broadly, an index that conflates the treatment signal with correlated but irrelevant variation, such as soil background reflectance in a tillage study or atmospheric aerosols in a region with high seasonal haze, can introduce bias that is difficult to detect. An index well suited for crop type mapping may therefore be poorly suited to estimating the effect of an intervention on vegetation health, even if both tasks rely on the same underlying imagery.

Plotting VI time series and spectral measurements early in an analysis can help identify which indices track the phenomenon of interest more effectively, and prior literature may provide context-specific approaches worth adapting.

\textbf{NDVI's limitations and alternatives.} NDVI is often a useful starting point for analysis, but its limitations are well documented. It saturates in high-biomass environments, losing sensitivity precisely where vegetation is densest. It can also conflate photosynthetically active vegetation with soil background reflectance, crop residue, weeds, and atmospheric aerosols (Huete 1988). Researchers have developed several alternative indices, each designed to address specific problems:

\begin{itemize}
\item \emph{Saturation in dense canopies:} The enhanced vegetation index (EVI; Huete et al. 2002) incorporates blue-band corrections and maintains sensitivity in high-biomass settings.
\item \emph{Soil background contamination:} The soil-adjusted vegetation index (SAVI; Huete 1988) applies an empirical correction factor to reduce the influence of soil reflectance in sparsely vegetated areas.
\item \emph{Green biomass and chlorophyll estimation}: NDVI's red band is relatively insensitive to chlorophyll content at higher biomass levels, limiting its usefulness in dense canopies. The green chlorophyll vegetation index (GCVI; Gitelson et al. 2003) addresses this by using the green and NIR bands rather than red and NIR, often providing a closer approximation of green leaf biomass and chlorophyll content. EVI can also improve performance in these settings by reducing saturation effects through its blue-band corrections. That said, performance remains dependent on context, and neither index outperforms NDVI consistently across all settings.
\item \emph{Spectrally similar crops:} Red-edge band indices exploit the steep reflectance transition between visible and NIR wavelengths to distinguish crops with similar broadband spectral profiles, such as wheat and barley (Ashourloo et al. 2022). Figure~\ref{fig:8-7} illustrates this challenge: Barley and wheat exhibit highly similar reflectance patterns across the visible bands and even into the red-edge region, with clearer separation emerging only at longer NIR wavelengths.
\end{itemize}

No single index is universally optimal. Calculating multiple indices and assessing their performance for the study area and application context is therefore good practice before selecting one for analysis.

\textbf{Common agricultural applications.} EO-derived variables support a wide range of agricultural applications relevant to GIE (Weiss et al., 2020). These include crop type and land cover mapping (Fritz et al. 2015, 2019); yield and biomass estimation from peak or aggregate VI values (Labus et al. 2002), including approaches that require no ground calibration, such as the scalable satellite-based crop yield mapper (SCYM; Lobell et al. 2015) and the Versatile Versatile Crop Yield Estimate; phenological monitoring of planting and harvest dates using smoothed VI time series (Diao 2020; Shen et al. 2022); field boundary delineation (Estes et al. 2022; Kerner et al. 2025); and detection of specific management practices, including irrigation (Deines et al. 2019a), cover cropping (Deines et al. 2023; Seifert et al. 2018), conservation tillage (Deines et al. 2019b), and soil and water management (Ali et al. 2020). Each application relies on a predictive model, whether simple or machine learning--based, and each introduces sources of uncertainty that can propagate into downstream GIE analyses.

\textbf{Smallholder contexts.} All predictive models perform worse in smallholder settings for several compounding reasons: higher cloud cover during tropical and subtropical growing seasons, small field sizes that increase mixed-pixel contamination, intercropping and agroforestry that blur spectral profiles, and highly variable management practices. A striking consequence is that models calibrated with crop cut data can underperform uncalibrated models in these settings, since noisy ground reference data may introduce more error than they correct (Lobell et al. 2020). This does not mean ground data are useless, but rather that their collection, quality, and representativeness require just as much scrutiny as the EO data.

\textbf{Crop masks.} Restricting analysis to relevant land areas using crop masks is generally expected to improve both classification accuracy and causal model precision, though the magnitude of these gains depends on the study context and mask quality. Crop masks determine which pixels are included in an analysis. Restricting the sample to cropland, or to a specific crop type, can help avoid contaminating estimates with spectral signals from forests, urban areas, or water bodies. This is particularly important when aggregating pixel-level values to produce summary statistics, such as a mean or median NDVI for a district or administrative unit, where non-crop pixels can dilute or distort the agricultural estimates, especially in landscapes where cropland is spatially fragmented or interspersed with other land cover types.

However, this logic only holds if the mask itself is accurate. The net benefit of masking depends on mask quality in the study region, whether classification errors are random or systematic, and, crucially, whether mask errors are correlated with the treatment or outcome of interest. A mask that systematically excludes certain field types or management practices can introduce bias of its own. Crop masks are modeled products and inevitably contain classification errors, so treating them as ground truth is a mistake. Table~\ref{tab:8-1} illustrates these tradeoffs for three commonly used products, although many alternatives exist, and researchers can also construct their own masks from raw imagery when off-the-shelf products are unavailable or poorly suited to the study context. The right choice will depend on the study region, crop of interest, and temporal requirements.

\begin{longtable}[]{@{}
  >{\centering\arraybackslash}p{(\linewidth - 6\tabcolsep) * \real{0.1944}}
  >{\centering\arraybackslash}p{(\linewidth - 6\tabcolsep) * \real{0.2178}}
  >{\centering\arraybackslash}p{(\linewidth - 6\tabcolsep) * \real{0.2792}}
  >{\centering\arraybackslash}p{(\linewidth - 6\tabcolsep) * \real{0.3085}}@{}}
\caption{Commonly used cropland mask products and their key tradeoffs for use in agricultural remote sensing applications.}\label{tab:8-1}\\
\toprule\noalign{}
\endfirsthead
\toprule\noalign{}
\endhead
\bottomrule\noalign{}
\endlastfoot
\textbf{Product} & \textbf{Spatial resolution} & \textbf{Temporal
coverage} & \textbf{Key tradeoff} \\
USDA Cropland Data Layer & 30 m & Annual (US only) & High classification
accuracy domestically, though performance varies across crop types and
regions; not globally available \\
WorldCereal (Van Tricht et al. 2023) & 10 m & 2021 baseline (global);
extensible via the Reference Data Module (RDM), a crowd-sourced ground
truth system that allows users to contribute local crop observations and
retrain the classifier for new years or regions & Distinguishes cereal
crop types at fine resolution, but baseline coverage is limited to 2021;
the RDM enables targeted updates where local reference data are
available \\
Dynamic World (Brown et al. 2022) & 10 m & 2015--present, near real-time
(global) & Enables near real-time, in-season mapping, though in-season
classifications are more difficult to validate reliably \\
\end{longtable}

Choosing among these products requires matching their strengths and limitations to study needs. WorldCereal's crop type specificity is particularly valuable for distinguishing cereals, but its baseline coverage constrains longitudinal analyses unless additional retraining data are available.

The Dynamic World product enables near real-time, in-season mapping, which is a genuine advantage for studies requiring contemporaneous land cover information. However, its validation poses a structural challenge. Because Dynamic World outputs a continuous feed of per-image predictions rather than a fixed annual map, users must make their own temporal aggregation choices, for example by assigning each pixel its most frequently observed class (the modal class) across a growing season or a calendar year. Brown et al. (2022) acknowledge that rigorous design-based accuracy assessment is difficult to apply to a product defined this way, since the set of pixels assigned to each land cover class changes depending on the aggregation strategy. Validation and performance metrics reported for Dynamic World therefore reflect a specific aggregation approach and may not transfer directly to a researcher's own derived product.

Independent comparative assessments further suggest that apparent differences in cropland classification across products often reflect definitional inconsistencies as much as genuine differences in accuracy. For example, Dynamic World includes fallow plots within its cropland class, while some validation datasets classify these areas as herbaceous vegetation, inflating apparent confusion rates in cross-product comparisons (Venter et al. 2022, Xu et al. 2024). For researchers, this means that disagreement between products does not necessarily indicate that one product is incorrect; it may instead reflect different, equally defensible, definitions of what counts as cropland.

These definitional inconsistencies and accuracy challenges are not unique to Dynamic World. Cropland accuracies for multiple global maps fall below 65\% in parts of Africa (Nabil et al. 2020), and consensus across widely used products can be remarkably low. Kerner et al. (2024) find that the 11 maps they evaluate unanimously classify fewer than 0.5\% of pixels as cropland across eight African countries, despite Food and Agriculture Organization (FAO) estimates indicating that a much larger share of land is used for agriculture. This suggests widespread disagreement among products over which pixels are agricultural, implying that for most agricultural pixels, at least one map is misclassifying them.

Regional accuracy assessments should therefore be reviewed carefully before committing to a product. Quality flags and confidence bands should also be examined where available. Where no clearly superior option exists, running analyses with multiple masks and assessing sensitivity to mask choice is likely preferable to treating any single product as definitive.

\textbf{Crop calendars.} A crop calendar is a structured record of the typical timing of key crop growth stages, including planting, emergence, flowering, and harvest, for a given crop and region. These calendars serve two main purposes in EO-based analyses.

\begin{figure}[htbp]
  \centering
  \includegraphics[width=\textwidth,height=0.75\textheight,keepaspectratio]{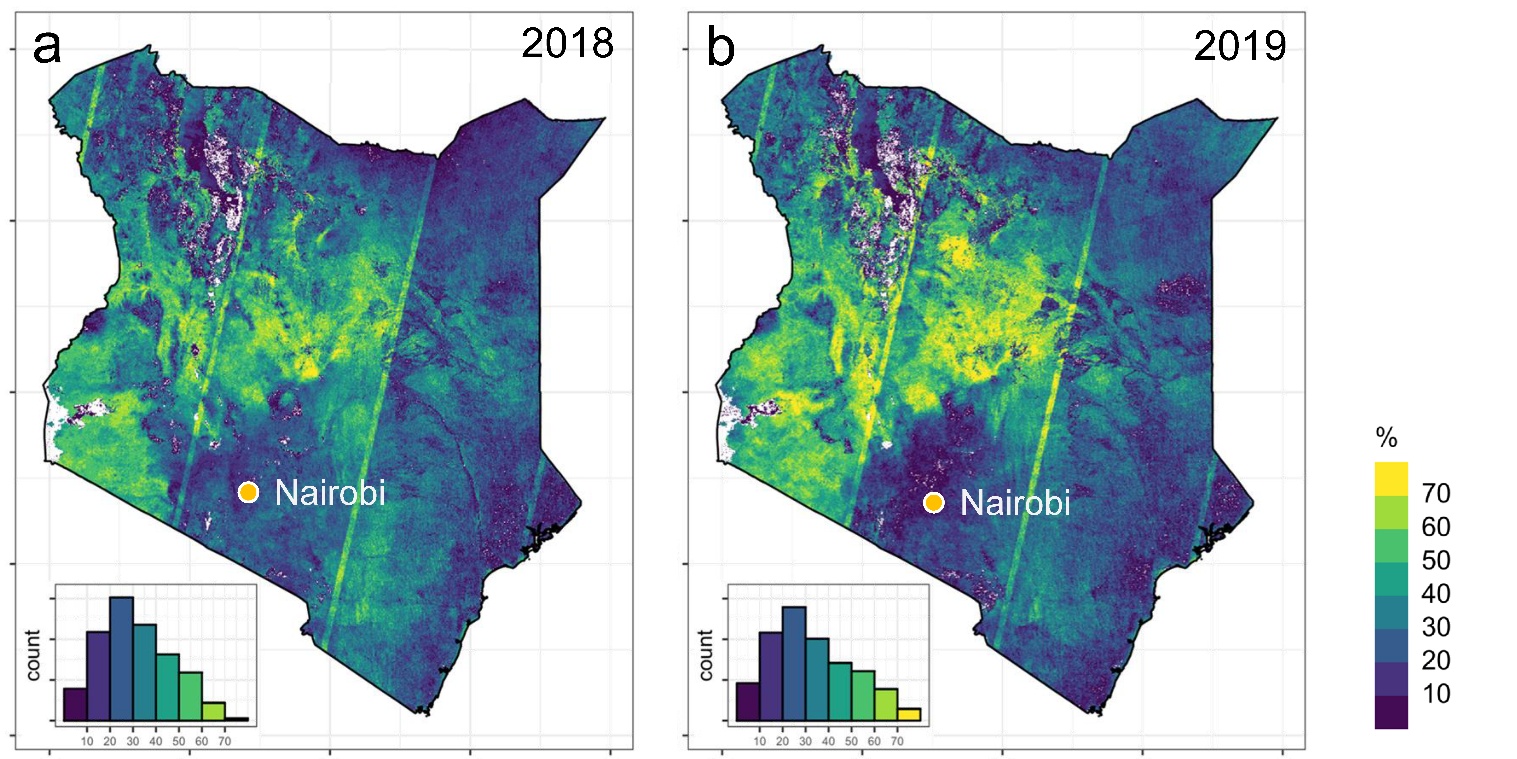}
  \caption{Cloud cover and usable imagery during the April--September rainy season in Kenya in 2018 and 2019. The percentage of Sentinel-2 observations usable for analysis, defined as pixels not classified as clouds or cloud shadows, varies substantially by location and year. From Benami et al. 2021.}
  \label{fig:8-9}
\end{figure}

First, in crop type classification, crop calendars help identify the periods when crops are most spectrally distinct. Peak growing season periods typically provide the greatest class separability. As Figure~\ref{fig:8-8} illustrates, barley, corn, and wheat are nearly indistinguishable in winter months but diverge substantially during the growing season. Nevertheless, it is worth investigating how well the growing season is represented in the imagery. This is particularly important in tropical and subtropical settings, where the rainy season coincides with the cropping season. Rainfall is the primary driver of agricultural production in most rainfed systems, but the same conditions that make rain agronomically critical also generate cloud cover that reduces the availability of usable imagery. In Kenya, for example, the proportion of usable Sentinel-2 observations varies substantially across locations and years (Benami et al. 2021), as shown in Figure~\ref{fig:8-9}. Western Kenya, one of the country's most productive agricultural regions, generally exceeds 50\% usable observations, though coverage remains highly variable across both space and time. A model trained or applied without accounting for this issue may rely on patchy or uneven growing season coverage, meaning that the imagery used during agronomically critical periods may be less representative than assumed.

Second, for causal analyses, crop calendars define the agronomically relevant window for linking weather or management variables to outcomes. Growing season rainfall typically has the strongest and most direct relationship to production, though pre-season conditions may also matter where they affect soil moisture, waterlogging, or input application decisions. These considerations are compounded in regions with multiple cropping seasons within a single year and substantial spatial variation in growing season timing. In parts of East and West Africa, some areas support two or even three distinct cropping cycles annually, while neighboring areas at different elevations or rainfall regimes follow just one season. Applying a single national crop calendar in these settings risks using the wrong time window for some observations. Where such misalignment occurs, the resulting measurement error may be spatially patterned and correlated with location or rainfall regime, making it difficult to diagnose without ground truth data on actual planting dates. Observed spatial patterns may therefore reflect genuine agroclimatic variation, calendar misspecification, or both.

\section{What is a shock? Incorporating rare events and extremes}

Researchers are often interested not only in measuring contemporaneous weather conditions, but also in assessing household and agricultural responses to extreme or rare events, commonly referred to as shocks. To define a shock as a deviation from expected conditions, researchers need both contemporaneous data to observe the event and historical data to establish a baseline against which it is judged rare or extreme. The choice of baseline is consequential: The reference window used (e.g., 10, 20, or 30 years) determines what qualifies as extreme, and under ongoing climate change, a fixed historical baseline may no longer reflect current conditions, potentially biasing shock measurement (Fischer and Knutti 2015). A rainfall deficit that looks extreme against a 10-year baseline may appear unremarkable against a 30-year one that includes historically drier periods, or vice versa.

Determining how large a deviation must be to constitute a shock, and how that deviation should be quantified, is not always straightforward. In many applications, the operational challenge is less whether a shock occurred than whether it should be considered extreme in a statistical, agronomic, or welfare-relevant sense. Misclassifying an event as extreme or non-extreme can lead researchers to misinterpret estimated coefficients, causing them to conflate gradual exposure effects with genuine shock effects, or to overlook the fact that rare events may affect different populations or environments in different ways.

Many empirical studies define shocks using threshold-based approaches, flagging a weather event as a shock if it exceeds a specific cutoff relative to a long-run mean or distribution. Thresholds can be defined in absolute terms, for example temperature above 40°C or rainfall below 10 mm per week, or relative terms, for example, more than two standard deviations above the historical mean or above the 90th percentile. In the latter case, the classification could operate as follows: If historical temperatures average 30°C with a standard deviation of 3°C, a two-standard-deviation threshold would classify temperatures above 36°C as shocks, whereas a 90th-percentile threshold might classify all temperatures above 35°C as extreme because only the hottest 10\% of historical observations exceed that value.

Absolute thresholds typically correspond to physiological or agronomic limits, whereas relative thresholds are defined in relation to typical local conditions, meaning that the same threshold may correspond to very different temperatures or rainfall across locations. Thresholds should therefore ideally be crop- and location-specific: A temperature threshold appropriate for maize during flowering may be too high or too low for rice because of differences in heat tolerance and growth stage sensitivity. For both types of approach, the operational choices made matter. Within relative threshold approaches, the choice of one versus two standard deviations determines how rare an event must be to qualify as a shock, while the median can be substituted for the mean as a more robust reference point when distributions are skewed. More generally, constructing separate indicators for above- and below-threshold events allows researchers to model asymmetric responses to excess and deficit conditions rather than assuming they are mirror images of one another.

Threshold-based approaches dominate largely for pragmatic reasons: They are simple to compute, easy to interpret, and map cleanly into regression specifications. Agronomically grounded variants such as growing degree days (GDDs) and killing degree days (KDDs) tend to offer a more principled approach by capturing beneficial heat accumulation and damaging heat stress within the same framework, as discussed in the Temperature section. However, these approaches require crop- and location-specific parameters that are not always available. Applying fixed agronomic thresholds also assumes that the underlying temperature data are well calibrated, which is not always the case. As discussed above, land surface temperature derived from remote sensing can diverge systematically from air temperature, and station-based records may carry their own biases. Where such biases exist, thresholds defined in absolute terms may misclassify events, making correction or recalibration of the underlying data necessary before thresholds can be meaningfully applied.

In many low-income settings, sparse station networks constrain researchers to standardized anomalies or deviations from historical baselines as a practical fallback. In addition, many widely used climate products are distributed as seasonal anomalies rather than as raw rainfall or temperature values, further encouraging anomaly-based shock definitions.

Despite their prevalence, however, threshold-based measures are often implemented without clear justification. Threshold selection is ultimately a modeling choice tied to the underlying causal mechanism, and even small adjustments to thresholds can substantially alter which events are classified as shocks and, in turn, the estimated effects. A related concern is that threshold-based methods assume comparability across space and time, which may not hold in practice. A one-standard-deviation rainfall deficit may be inconsequential in one setting and devastating in another depending on crop type, soil water storage, and evapotranspiration demand. To improve comparability, some researchers use z-scores (the deviation from the historical mean divided by the standard deviation) or the coefficient of variation (CV; the standard deviation divided by the mean, useful when comparing locations with very different average rainfall levels). These measures reduce sensitivity to distributional differences across locations, but they capture statistical extremeness rather than agronomic or welfare relevance. A statistically large deficit, for example, may have little effect if it falls outside the period of peak water demand. In practice, researchers typically apply a cutoff to these measures, reintroducing the threshold selection problem.

Regardless of the approach used, documenting and, where possible, validating threshold choices against observed crop or welfare responses is important for transparency and reproducibility.

\subsection{Droughts}

Drought is a term in common circulation, and most readers will feel they recognize it when they encounter it. That intuitive familiarity is somewhat deceptive, however, because drought is inherently complex and multifaceted. Scholars have long distinguished between conceptual definitions (e.g., a prolonged dry period) and operational definitions specifying onset, severity, frequency, and duration. In brief, meteorological drought refers to precipitation shortfalls; agricultural drought to soil moisture deficits that impair crop growth; hydrological drought to sustained declines in river, reservoir, or groundwater availability; socioeconomic drought to the resulting human and economic consequences; and ecological drought to persistent moisture deficits affecting ecosystem function and biodiversity (AghaKouchak et al. 2015). Because these definitions capture different dimensions of drought, researchers need to clearly justify which drought metric best matches the phenomenon of interest.

Socioeconomic researchers often use precipitation anomalies as a convenient proxy for drought, since they can be derived readily from EO rainfall products. Yet, precipitation anomalous remain a coarse approximation of drought. As discussed in the Precipitation section and illustrated in Figure~\ref{fig:8-3}, precipitation estimates can be highly skewed, with occasional extreme events shaping its distribution, and what constitutes a meaningful shortfall varies substantially across contexts. Several alternative indicators have therefore been developed to address these limitations, generally by incorporating water demand alongside supply.

The Palmer drought severity index (PDSI) incorporates precipitation and temperature into a water supply-and-demand model of soil moisture (Palmer 1965). Its long smoothing window makes it well suited to multi-month or multi-year drought characterization but slow to capture short-term onset (Dai 2011). The standardized precipitation evapotranspiration index (SPEI) accounts for both rainfall and potential evapotranspiration, enabling characterization of drought onset, duration, and severity across multiple time scales (Vicente-Serrano et al. 2010; Beguería et al. 2014). The water requirement satisfaction index (WRSI), used operationally by the Famine Early Warning Systems Network (FEWS NET) across the Sahel and Horn of Africa, explicitly models whether crop water requirements are met over a growing season, making it more directly actionable for GIE than climatological anomaly measures (Verdin and Klaver 2002; Senay and Verdin 2003). Soil moisture indices derived from remote sensing and hydrological models can also inform water stress assessment but, as noted in the Mismeasurement section, they frequently rely on coarse-resolution or modeled products that introduce uncertainty in heterogeneous or data-sparse regions (McNally et al. 2015). Mishra and Singh (2010) provide a systematic review of drought indices and their strengths and limitations for varied socioeconomic applications.

Overall, these indices are complementary rather than interchangeable: PDSI is better suited to characterizing longer-run moisture balance; SPEI is readily applicable to short- to medium-term drought given its relatively modest data input requirements; WRSI is explicitly calendar- and crop-specific; and soil moisture indices target plant-available water most directly but, given their high data demands, may be estimated with high data uncertainty.

Growing interest in moving beyond single-index approaches has produced integrated drought measures that combine remotely sensed precipitation and temperature with vegetation indices such as NDVI anomalies, thereby exploiting the complementary strengths of multiple inputs. The vegetation drought response index (VegDRI), for example, combines climate, soil, and satellite data on this basis (Brown et al. 2008).

Such integration reduces reliance on any single assumption but also compounds uncertainty across inputs, potentially making the resulting index more difficult to validate or interpret. Its primary value therefore lies less in the precision of its outputs than in offering confidence in the direction of detection When precipitation deficits, soil moisture anomalies, and vegetation stress indicators independently converge, confidence in shock identification increases even if each individual underlying indicator remains uncertain (Enenkel et al. 2019; Richaud et al. 2024).

\subsection{Floods}

Flood measurement poses its own definitional challenges. Unlike temperature or precipitation, which are continuous variables that take a value for every location and time, floods are discrete events whose occurrence and extent depend on the interaction of weather with local hydrology, terrain, and infrastructure (Mishra et al., 2022). Defining when a flood has occurred, and where, therefore requires choices that go beyond the underlying physical data. Researchers have drawn on two broad types of flood records: event-based archives compiled from media reports, government sources, and field observations; and satellite-derived inundation products that estimate surface water extent from sensor measurements.

A well-known event-based resource is the Dartmouth Flood Observatory (DFO) archive, which compiles flood event records from media and government reports dating back to 1985. Unlike satellite-derived inundation products, the DFO provides rich ancillary information for each event, including primary cause, affected area, river names, duration, reported deaths and displaced persons, damage estimates in US dollars, and a severity classification (Langlois et al. 2024). The archive has been used in economics research to study the effects of floods on urban economic activity and development (e.g., Kocornik-Mina et al. 2020), among other applications. However, these advantages come with significant coverage limitations. Because the archive relies on opportunistic reporting, coverage is uneven across time and space and systematically favors large, high-visibility floods in more connected regions. As a result, detailed inundation maps exist predominantly for a subset of floods, smaller and more remote flood events are systematically underrepresented, and updates to the archive are inconsistent (Langlois et al. 2024). Analyses that treat the DFO as a comprehensive record of flood occurrence therefore risk conflating true variation in flood exposure with variation in reporting intensity.

Satellite-derived inundation products address some of these limitations by detecting surface water directly from imagery rather than relying on reported events. Tellman et al. (2021), for example, constructed a global flood database from MODIS imagery, mapping 913 flood events between 2000 and 2018 at 250 m resolution and enabling researchers to quantify the proportion of a population or agricultural area exposed to inundation. This and similar products have been used in economics research to identify flood exposure and estimate welfare and development effects (Chen et al. 2017; Collalti 2024; Patel 2023; Michler et al. 2026).

However, satellite-based products also carry important limitations. Cloud cover frequently obscures optical sensors during the rainfall events that cause floods, precisely when detection is most needed. MODIS's 250 m resolution makes it poorly suited for mapping floods in urban areas or in contexts with heterogeneous terrain, while its revisit frequency can miss short-duration inundation events entirely. Synthetic aperture radar (SAR) sensors such as Sentinel-1 can observe through cloud cover and at night, making them increasingly preferred for operational flood mapping (Robert Brakenridge et al. 2012). However, the Sentinel-1 archive began only in 2014, limiting its usefulness for longer-run retrospective analyses. Patel (2023) and Giezendanner et al. (2023) both address these limitations by fusing optical and radar satellite data using deep learning, enabling more complete historical inundation records than either sensor provides alone.

A third approach uses physics-based hydrological modeling to simulate flood potential from antecedent soil moisture, terrain, and rainfall inputs, ranging from simplified one-dimensional mass balance models to full shallow-water equation models (Bates et al. 2010; Grimaldi et al. 2019). These methods can extend flood detection to locations and time periods without satellite observations and form the basis of operational early warning systems such as the Global Flood Awareness System (GloFAS; Harrigan et al. 2020). However, they introduce their own assumptions and uncertainty and require careful calibration to local topographic and hydrological conditions for which information may be unavailable in data-sparse settings (Bates 2023).

Regardless of the measurement approach used, defining what constitutes a flood shock, rather than seasonal or permanent water, requires a threshold. Remotely sensed flood extents are typically compared against baselines of permanent surface water, such as the Joint Research Centre (JRC) Global Surface Water Explorer dataset, to isolate anomalous inundation. Z-score thresholds can then be applied to the proportion of a spatial unit classified as flooded in order to define extremes. As with drought and temperature measures discussed above, however, these thresholds remain context dependent. Flood impacts depend not only on inundation extent, but also on depth, duration, flow velocity, soil infiltration capacity, infrastructure, and the timing of an event within the agricultural calendar. Standing water may benefit paddy rice during land preparation but devastate the crop during grain filling (Wang et al. 2022). Michler et al. (2026) reinforce the importance of precise flood characterization for identifying relevant treatment effects, as the assessment of submergence-tolerant rice varieties in Bangladesh hinges on clarifying the exact timing and severity of floods (e.g., identifying the ``Goldilocks flood'').

A separate challenge is that inundation at a given location is governed by basin hydrology rather than local conditions alone. Precipitation falling upstream can drive flood exposure at a downstream location days or even weeks later (Guiteras et al. 2015), meaning that even well-measured local inundation may reflect a lagged and spatially displaced consequence of distant weather events. This has important implications for GIE study design: A unit downstream from a flood may experience delayed hydrological effects (waterlogging, elevated soil moisture, and infrastructure damage) that are not captured in satellite inundation maps at the time of the event, causing it to be coded as a control unit when it has in fact been affected. This kind of latent exposure can attenuate estimated treatment effects and is difficult to detect without detailed hydrological knowledge of the study area.

As an alternative to purely physical thresholds, Moore and Obradovich (2020) propose a socially grounded definition of shocks based on remarkability, defined as the perceived unusualness of events as measured through deviations in local media reporting. In operational settings, similar logic underlies efforts to detect shocks for humanitarian response: Jongman et al. (2015), for example, combine satellite-derived flood extent with flood-related Twitter activity to identify events warranting rapid response. Although these approaches differ, they share the view that linking observational data to human experience is important, as even physically comparable floods can have very different impacts depending on local preparedness, societal memory of prior events, and whether they exceed the capacity of existing infrastructure (Di Baldassarre et al. 2015).

\subsection{Synthesis}

The less explicitly a shock measure is defined, the harder it becomes to interpret what exposure it captures, identify the mechanisms through which it operates, and translate findings into actionable policy guidance. Measures such as z-scores, percentiles, and growing degree day interactions are not inherently problematic, but without clear documentation and justification of the reference period, distributional assumptions, and threshold choices that underpin them, the resulting effect estimates are difficult to audit, replicate, or transfer across contexts.

\section{Issues of mismeasurement (and misuse)}

Throughout this chapter, we have discussed the numerous (and growing) sources of data on meteorological and terrestrial phenomena available for geospatial impact evaluations. Yet this plethora of data sources should not obscure a fundamental point: Satellite-based estimates are not themselves ``truth.'' They are modeled estimates shaped by sensor degradation, calibration drift, georegistration error, and modeling assumptions. Even the underlying observational archives carry systematic gaps: Remelgado et al. (2023) show that the global Landsat record is spatially and temporally uneven, and that these inconsistencies propagate into widely used time-series products in ways that can bias environmental change analyses. Some of these errors are classical in that they introduce noise that obscures the relevant signal without systematically biasing estimates. Others are non-classical and can generate systematic bias if used uncritically.

What makes this more than a conventional data quality problem is that EO measurement errors are frequently patterned in ways that are spatially and socioeconomically correlated with the very interventions and populations researchers seek to study. From an econometric standpoint, classical measurement error in an independent variable typically causes attenuation bias, pushing estimated treatment effects toward zero. Non-classical error is more problematic: It can exaggerate effects, reverse their sign, or induce spurious spatial correlations that mimic program impacts. If a rainfall product systematically underestimates precipitation in mountainous areas but not in plains, for example, an evaluation of irrigation or drought relief programs may falsely attribute yield differences to the intervention rather than to spatially patterned measurement error in the rainfall product. Josephson et al. (2025) provide direct evidence of this problem: The sign and magnitude of estimated rainfall--yield relationships vary substantially across remotely sensed products and even within a single product across different contexts, meaning researchers working with different data sources may reach contradictory conclusions about the same underlying reality. Sohnesen (2020) documents a similar pattern for drought, finding that different drought indicators yield substantively different estimates of Ethiopia\textquotesingle s 2015 drought impacts on household consumption.

Although these concerns have been discussed most extensively in the context of satellite-derived deforestation data, the same logic applies to remotely sensed drought and agricultural shocks. Where measurement errors are systematically related to the true underlying conditions (e.g., dense forest cover or mountainous terrain), the resulting errors may not simply attenuate estimated effects but distort them in unpredictable ways (Alix-García and Millimet 2023; García and Heilmayr 2024).

Before turning to potential remedies, it is worth identifying several common misuses that motivate this discussion. These include: (1) using a global or regional EO product in a local context without first validating it against ground observations from that setting; (2) selecting among multiple available products based on which produces the most statistically significant or theoretically expected coefficient, a form of specification search that exploits product-level measurement error rather than addressing it; (3) treating Level 4 modeled outputs (e.g., estimated crop yield or gross primary productivity; see below) as equivalent to direct observations without accounting for the embedded modeling uncertainty; and (4) assuming that higher spatial resolution implies greater accuracy, even though fine-grained gridded products in data-sparse regions may simply interpolate noise at finer scales. Awareness of these failure modes motivates the framework that follows.

Even in an era of rapidly expanding and improving remote sensing data, ground reference data are and will continue to be essential for four distinct tasks: \textbf{validation}, \textbf{calibration}, \textbf{debiasing}, and \textbf{model building}. Although related, each addresses measurement error differently and is appropriate for different error structures. \textbf{Table 8.2A} provides a consolidated reference for these concepts and their relationship to error types; we elaborate on each below.

\begin{itemize}
\item \emph{Validation} assesses alignment between remotely sensed values and ground truth data. What constitutes acceptable performance depends on the variable's role in the GIE: A variable central to causal identification demands a higher standard than a covariate included for statistical control.
\item \emph{Calibration} applies relatively simple adjustments, such as linear rescaling, to address classical mismeasurement error, including uniform offsets (where values are consistently too high or too low) or scaling differences (where variation is systematically exaggerated or compressed).
\item \emph{Debiasing} is warranted when error is non-classical and varies with context: for example, when rainfall is undermeasured in mountainous terrain or NDVI values saturate in dense forests. In such cases, a single global rescaling is insufficient. Some approaches combine ground reference data with contextual predictors to estimate context-dependent corrections (Alix-Garcia and Millimet 2022; Proctor et al. 2023), possibly complemented by adjustments to the prediction model itself that reduce errors correlated with the treatment variable (Gordon et al. 2026).If differential error is ignored, it can induce spurious correlations and bias treatment effect estimates.
\item \emph{Model building} involves constructing predictive models that relate EO inputs to a target variable using ground reference data. It is appropriate when these relationships are nonlinear or context dependent, or when EO measures serve only as indirect proxies for the construct of interest. Model-building approaches typically involve machine learning methods trained on ground reference data, EO inputs, and control variables, and evaluated via cross-validation and out-of-sample prediction (Burke et al. 2021; Ratledge et al. 2022).
\end{itemize}

The continued need for ground reference data reflects two persistent issues: the relationships that remote sensing models learn can shift across regions and over time (nonstationarity), and remotely sensed observations often serve as imperfect proxies for the target variable of interest. The subsections below summarize task-specific guidance for the core components of model building.

\subsection{Level designations.}

Remote sensing products often carry processing-level designations indicating the extent to which raw sensor data have been corrected, transformed, or modeled. Level 1 products contain calibrated radiance; Level 2 products provide atmospherically corrected geophysical variables (e.g., surface reflectance, vegetation indices, and land surface temperature) while retaining the sensor's original spatial resolution; Level 3 products aggregate data to a uniform grid; and Level 4 products contain modeled outputs such as gross primary productivity, evapotranspiration, or soil moisture (Justice et al. 1998). This convention is used across major sensor families, including MODIS, Landsat, VIIRS, and Sentinel. For most practitioners, Level 2 or higher products are generally most appropriate. However, Level 4 products embed additional modeling assumptions and may carry uncertainty that propagates into downstream estimates in nontransparent ways. Users should therefore consult quality flag documentation and validate outputs against ground data where feasible. Researchers should also be attentive to version changes within a processing level, such as the transition from MODIS Collection 5 to Collection 6, which can introduce temporal discontinuities that mimic real-world trends if series are spliced without adjustment.

\subsection{Ground data collection.}

Rapid improvements in remote sensing data contrast with a persistent shortfall in reliable ground data, especially in smallholder settings (Burke et al., 2021). Even small amounts of high-quality training data, on the order of 30--50 samples, can significantly improve model performance in constrained domains (Burke et al. 2021), while a few hundred samples may suffice for bias correction via multiple imputation (Proctor et al. 2023). Broader, multi-context studies typically require substantially larger samples. Effective ground data collection hinges on both sampling strategy and labeling quality. Sampling design determines how many observations are collected and where, often using remote sensing to stratify by land cover, elevation, or topography. Labelling and metadata should capture target variables, management practices, plot boundaries, and crop phenology timing. When proxies are used in place of direct measurement, such as biomass for yield or nighttime lights for economic activity (Donaldson and Storeygard 2016), the proxy relationship introduces additional uncertainty that should be explicitly accounted for in GIE estimates.

\subsection{Model training and loss functions.}

Standard least-squares regression models optimize average fit, but this objective is not always aligned with the goals of GIEs. When evaluation targets, such as vulnerable households, extreme weather events, or low-performing farms, are concentrated in the tails of a distribution, models should be trained and evaluated accordingly. In index insurance applications, for example, loss functions can be designed to minimize missed payouts during severe production shortfalls (Benami et al. 2021). Similarly, quintile-optimized training can improve performance at the distributional extremes relative to standard regression approaches (Ratledge et al. 2022). Overall, loss functions should reflect the evaluation's inferential objectives rather than defaulting to computational convenience.

\subsection{Identifying and addressing errors}

Diagnosing the structure of measurement error is a prerequisite for selecting an effective correction strategy. Jain (2020) demonstrates that EO models systematically miss low-intensity and small-scale events, including localized deforestation and the lowest-yielding fields, potentially biasing impact estimates when treatment or outcomes are concentrated in those domains. Josephson et al. (2025) show that differential mismeasurement across rainfall products can distort causal estimates in both magnitude and direction. Diagnostic approaches include tracing upstream classification errors through to downstream production estimates (Estes et al. 2018) and stress-testing results across plausible ranges of values for key remotely sensed variables (Fowlie et al. 2019). Table~\ref{tab:8-2} summarizes three major error types, their diagnostic signatures, and corresponding correction approaches, drawing on the classification proposed in Proctor et al. (2023).

\begin{longtable}[]{@{}
  >{\raggedright\arraybackslash}p{(\linewidth - 8\tabcolsep) * \real{0.1608}}
  >{\raggedright\arraybackslash}p{(\linewidth - 8\tabcolsep) * \real{0.1755}}
  >{\raggedright\arraybackslash}p{(\linewidth - 8\tabcolsep) * \real{0.2094}}
  >{\raggedright\arraybackslash}p{(\linewidth - 8\tabcolsep) * \real{0.2153}}
  >{\raggedright\arraybackslash}p{(\linewidth - 8\tabcolsep) * \real{0.2389}}@{}}
\caption{Error types, diagnostic tasks, and correction approaches for EO data in GIE. These tasks are not mutually exclusive: Researchers will typically begin with validation to diagnose the error type, then apply calibration or debiasing before using corrected or model-derived outputs in estimation. Adapted from Proctor et al. (2023).}\label{tab:8-2}\\
\toprule\noalign{}
\endfirsthead
\toprule\noalign{}
\endhead
\bottomrule\noalign{}
\endlastfoot
\textbf{Error type} & \textbf{Description} & \textbf{Consequence for
GIE} & \textbf{Diagnostic task \& methods} & \textbf{Correction
approaches} \\
Classical (random) & Random noise, uncorrelated with other variables &
Attenuates treatment effects toward zero; reduces statistical power &
Validation using root mean square deviation (RMSE), correlation, and
bias statistics; comparison against held-out ground data &
Higher-quality EO products; averaging across time or sensors;
calibration via linear rescaling to station or field measurements \\
Mean-reverting & Extremes systematically underpredicted; distribution
compressed & Underestimates impacts of shocks or interventions
concentrated at distributional extremes & Validation against
distributional tails; comparison of variance across products &
Alternative loss functions; quantile-optimized training; targeted
sampling at extremes \\
Differential / non-classical & Error magnitude or direction varies
systematically with context (topography, land cover, season, etc.) &
Generates spurious correlations; bias can vary in magnitude or even
reverse sign across subgroups; pooled corrections may become invalid &
Stratified validation by context; product triangulation & Debiasing with
contextual predictors; multiple imputation; targeted ground collection
in high-error contexts \\
Complex / indirect proxy & Target variable inferred from EO signals via
nonlinear or indirect relationships & Compounds the above risks; model
uncertainty propagates into GIE estimates & Cross-validation; ensemble
disagreement used as an uncertainty indicator & Machine learning;
cross-validation; ensemble methods; alternative loss functions \\
\end{longtable}

\section{Good and better practices}

The recommendations in this section follow the framework of good and better practices outlined by Olofsson et al. (2014). In this framework, good and better are distinct from best practices. Good practices recognize the wide variety of research contexts, data constraints, and analytical objectives that researchers face, while better practices acknowledge that methods are always improving and that what is optimal in one setting or at one time may be inappropriate or infeasible in another. The aim is therefore to set a rigorous but flexible standard that can be applied across a wide range of contexts and research settings.

These recommendations follow a clear priority sequence: Identify the mechanism first, then match data to that mechanism before evaluating product quality. A methodologically sophisticated product that fails to capture the relevant exposure is not an improvement over a simpler one that does. Good practice begins with taking this principles seriously, and the questions in Figure~\ref{fig:8-3} are organized to reinforce that sequence.

\textbf{Know your data before you use it.} Strong familiarity with any data product should precede its use in analysis. This means reading the documentation, understanding what the product is designed to measure, how it was constructed, its known limitations, and the contexts in which it has been validated. Documentation alone, however, is by no means sufficient. It conveys intended purpose but rarely anticipates the context-specific failures that matter in applied work. Researchers should therefore treat documentation as a starting point and explore the data directly, examining distributions, checking for outliers or discontinuities, and comparing values against local knowledge or alternative data sources. As EO products are frequently revised, good practice also requires recording the product name, version number, processing level, and release date to ensure your results remain reproducible and comparable over time.

\textbf{Match the data to the question.} Before selecting a data product, the single most important question a researcher should ask is: What is the mechanism through which weather affects the outcome I am studying, and does this product capture that mechanism at the relevant spatial and temporal scale? Answering this requires clarity about the biophysical or economic theory underlying the analysis (e.g., whether the relevant exposure is a seasonal deficit, a threshold exceedance, a short-duration extreme event, or a long-run trend), and about whether the product's resolution, period of record, and geographic coverage are adequate for the research context. The following questions can help guide product selection:

\begin{enumerate} \def\labelenumi{\arabic{enumi}.}
\item What is the spatial and temporal resolution of the product, and is it appropriate for the scale at which the outcome of interest varies?
\item What is the data generating process? How is the product constructed, calibrated, and updated?
\item What is the most appropriate way to quantify the exposure? Should it be aggregated daily, monthly, or seasonally? Are absolute levels or deviations from baseline more substantively relevant?
\item What sources of measurement error exist, and are they classical, non-classical, or differential? What steps has the producing organization taken to address known issues, and what validation has been conducted for the region and application of interest? What performance metrics are most appropriate, and do they reflect the demands of your intended use case (e.g., overall fit versus subgroup accuracy)?
\item Are there potential sources of endogeneity or systematic error that could bias inference or mislead policy guidance? For example, do patterns in data coverage systematically correlate with outcomes of interest?
\end{enumerate}

\textbf{Test robustness across products.} Because product choice can materially affect results, researchers should conduct robustness checks using alternative data products wherever feasible (Michler et al. 2022; Josephson et al. 2025). Convergent findings across distinct sources strengthen credibility, while divergent findings reveal product sensitivity that is itself informative and should be reported rather than concealed. Including an appendix comparing results across two or three products is a low-cost investment in transparency that can substantially increase confidence in reported findings. Burke and Tanutama (2019), for example, analyze the macroeconomic impacts of temperature using three different products, illustrating both the robustness of core findings and the variance in estimates that can arise from product choice.

\textbf{Consider alternative metrics alongside alternative products.} Robustness depends not only on which product is used, but also on how the weather variable is defined. Different metrics (e.g., mean temperature versus growing degree days, total rainfall versus rainfall above a threshold, or actual ET versus PET) can capture meaningfully different aspects of weather exposure and may yield different results even when derived from the same underlying product. Better practice therefore involves justifying metric choices with reference to the hypothesized mechanism, testing sensitivity to reasonable alternative definitions, and clearly documenting shock definitions and temporal windows, particularly for analyses involving extremes or rare events. Liu et al. (2020) apply multiple metrics to compare how different global fire emissions inventories represent the same fire events, illustrating how choices about data sources, inventory construction, and measurement approach can substantially affect downstream inference even when the underlying phenomenon being measured is ostensibly the same.

\textbf{Be transparent and reproducible.} Data and code sharing, clear documentation of processing steps, and explicit reporting of modeling decisions are foundational to credible empirical research, particularly in applied economics and social science, where policy recommendations often rest on findings that others cannot easily verify or replicate. This is especially important for EO-based analyses, where the transformation from raw data to analysis-ready variables involves many choices that are invisible to readers unless explicitly documented.

Weather and EO-derived data products have become substantially more accessible in recent years, with a rapidly expanding range of global, high-resolution, and near real-time products spanning weather, vegetation, and many other domains. But the core challenge this chapter has addressed--- that seemingly objective weather data are in fact constructed quantities whose properties depend on how they are generated---is not resolved by the expansion of available products. Researchers who bring the same critical scrutiny to their input data that they bring to their econometric identification strategies will produce more credible and cumulative evidence on the consequences of weather for human welfare.

\begin{chapbox} \textbf{Box 8.3: Questions to guide the choice of weather and land cover monitoring products}

\qcat{Familiarity with data generating process(es) relevant to your question(s)}
\begin{qlist}
\item What phenomenon are you studying (e.g., drought, yield, or yield variability) and how is it operationalized?
\item Have you explored your chosen dataset and reviewed the documentation to understand how the product has been generated? In particular:
  \begin{qlist}
  \item Does the product footprint match the scale of the question being addressed?
  \item What quality flags are provided?
  \item Are there important offline periods or gaps in coverage?
  \item Are there known georegistration errors, especially over time?
  \item Is there evidence of sensor degradation or drift?
  \item Have there been important version changes?
  \item Is validation or performance assessment documented across time, space, or different classes?
  \end{qlist}
\item Are ground reference datasets available to \textbf{cross-check} assumptions or \textbf{stress-test} results?
\item Which performance metrics are most appropriate for the intended applications (e.g., R²/overall variability versus subgroup fit, or asymmetric loss functions)?
\item How will uncertainty be assessed (e.g., confidence intervals or ensemble approaches)?
\item Can multiple datasets be compared using a convergence of evidence approach?
\item \textbf{What are the most likely sources of systematic error that could bias inference or mislead policy guidance?}
\end{qlist}

\qcat{Weather monitoring (Section 2)}
\begin{qlist}
\item What temporal resolution is required (e.g., short-lived extreme events versus seasonal trends)?
\item Which weather variable is most relevant, and where/how is it measured? (e.g., land surface versus air temperature at 2 m height)
\item Are ground station observations available, reliable, and representative of your study area, either as inputs to the selected product or for cross-checking product performance?
\end{qlist}

\qcat{Land cover/vegetation monitoring (Section 3)}
\begin{qlist}
\item What spatial and temporal scale is needed to detect the changes of interest, and which sensors and/or data processing algorithms are suitable to recover those changes?
\item How spectrally distinguishable are the land cover/land use classes you seek to evaluate, and in which spectral bands?
\item How might cloud cover and vegetation density affect observations (e.g., cloud mask quality, signal saturation, or VI performance)?
\item Are there terrain or landscape features that may complicate analysis, such as:
  \begin{qlist}
  \item variation/heterogeneity of soil or land cover types within your study area
  \item mountains, clouds and cloud shadows, or volcanic activity
  \end{qlist}
\end{qlist}

\qcat{Integration with Household/survey data (See Chapter 6)}
\begin{qlist}
\item Are field-level geospatial data available (e.g., polygons or points)?
\item Will mismatches in spatial resolution, privacy-related coordinate displacement (``jittering''), or geolocation aggregation complicate data integration''
\end{qlist}
\end{chapbox}

\section*{Exercises}

The exercises below use the Google Earth Engine (GEE) Code Editor and publicly available datasets. No prior coding experience is required. \href{https://courses.spatialthoughts.com/gee-sign-up.html}{\ul{This workflow}} from Spatial Thoughts provides step-by-step instructions for registering a GEE account. All exercise materials can be found in the ``Chapter8\_Exercise\_Materials'' folder.~

\href{https://docs.google.com/document/d/1Gyl39MYcHiO45Lx7_M_fLml9vLMpL3V5rF8sou-9Hq4/edit?usp=sharing}{\ul{Exercise 1}} -- Analyzing spectral signatures and time series using the US Cropland Data Layer

\href{https://docs.google.com/document/d/1rFymLKG3cahLXvn342E8XtnbbstQxDsKHZi3H7asv6k/edit?usp=sharing}{\ul{Exercise 2}} -- Analyzing spectral signatures and time series for crops in Morocco using WorldCereal

\href{https://docs.google.com/document/d/1vosM1TajsiiP8GyB-cpNnMtNZb9aVmmoqVPz6emudRE/edit?usp=sharing}{\ul{Exercise 3}} -- Comparing precipitation datasets

\section*{References}

Abadie, A., Athey, S., Imbens, G. W., \& Wooldridge, J. M. (2020). Sampling-based versus design-based uncertainty in regression analysis. \emph{Econometrica, 88}(1): 265-96.

Abatzoglou, J. T. (2013). Development of gridded surface meteorological data for ecological applications and modelling. \emph{International Journal of Climatology, 33}: 121-31.

Abatzoglou, J. T., Dobrowski, S. Z., Parks, S. A., \& Hegewisch, K. C. (2018). TerraClimate, a high-resolution global dataset of monthly climate and climatic water balance from 1958--2015. \emph{Scientific data}, \emph{5}(1), 170191.

Allen, R. G., Pereira, L. S., Raes, D., and Smith, M. (1998). Crop evapotranspiration: Guidelines for computing crop water requirements. FAO Irrigation and Drainage Paper No. 56. Food and Agriculture Organization of the United Nations, Rome.

Alix-García, J., \& Millimet, D. L. (2023). Remotely incorrect? Accounting for nonclassical measurement error in satellite data on deforestation. \emph{Journal of the Association of Environmental and Resource Economists, 10}(5), 1335--1367. \url{https://doi.org/10.1086/723723}

AghaKouchak, A., Mehran, A., Norouzi, H., \& Behrangi, A. (2012). Systematic and random error components in satellite precipitation data sets. \emph{Geophysical Research Letters}, \emph{39}(9).

AghaKouchak, A., Farahmand, A., Melton, F. S., Teixeira, J., Anderson, M. C., Wardlow, B. D., \& Hain, C. R. (2015). Remote sensing of drought: Progress, challenges and opportunities. \emph{Reviews of Geophysics, 53}(2): 452-80.

Ali, D. A., Deininger, K., \& Monchuk, D. (2020). Using satellite imagery to assess impacts of soil and water conservation measures: Evidence from Ethiopia's Tana-Beles watershed. \emph{Ecological Economics, 169}: 106512.

Allen, R. G., Pereira, L. S., Raes, D., and Smith, M. (1998). Crop evapotranspiration: Guidelines for computing crop water requirements. FAO Irrigation and Drainage Paper No. 56. Food and Agriculture Organization of the United Nations, Rome.~

Ashouri, H. (2015). Persiann-cdr: Daily precipitation climate data record from multisatellite observations. \emph{Bulletin of the American Meteorological Society, 96}(1): 69-83.

Ashourloo, D., Nematollahi, H., Huete, A., Aghighi, H., Azadbakht, M., Shahrabi, H. S., \& Goodarzdashti, S. (2022). A new phenology-based method for mapping wheat and barley using time-series of Sentinel-2 images. \emph{Remote Sensing of Environment, 280}: 113206.

Auffhammer, M., Hsiang, S. M., Schlenker, W., \& Sobel, A. (2013). Using weather data and climate model output in economic analyses of climate change. \emph{Review of Environmental Economics and Policy}.

Babaeian, E., et al. (2019). Ground, proximal, and satellite remote sensing of soil moisture. \emph{Reviews of Geophysics, 57}(2), 530--616. \url{https://doi.org/10.1029/2018RG000618}~

Bannari, A., Morin, D., Bonn, F., \& Huete, A. (1995). A review of vegetation indices. \emph{Remote Sensing Reviews, 13}(1-2): 95-120.

Barnett, T. P., Adam, J. C., \& Lettenmaier, D. P. (2005). Potential impacts of a warming climate on water availability in snow-dominated regions. \emph{Nature}, \emph{438}(7066), 303-309.

Bates, P. D., Horritt, M. S., \& Fewtrell, T. J. (2010). A simple inertial formulation of the shallow water equations for efficient two-dimensional flood inundation modelling. \emph{Journal of hydrology}, \emph{387}(1-2), 33-45.

Bates, P. (2023). Fundamental limits to flood inundation modelling. \emph{Nature Water}, \emph{1}(7), 566-567.

Beck, H.E., Vergopolan, N., Pan, M., Levizzani, V., Van Dijk, A.I., Weedon, G.P., Brocca, L., Pappenberger, F., Huffman, G.J. and Wood, E.F., 2017. Global-scale evaluation of 22 precipitation datasets using gauge observations and hydrological modeling. \emph{Hydrology and Earth System Sciences}, \emph{21}(12), pp.6201-6217.

Becker, A., Finger, P., Meyer-Christoffer, A., Rudolf, B., Schamm, K., Schneider, U., \& Ziese, M. (2013). A description of the global land-surface precipitation data products of the Global Precipitation Climatology Centre with sample applications including centennial analysis from 1901--present. \emph{Earth System Science Data, 5}: 71-99.

Beguería, S., Vicente-Serrano, S. M., Reig, F., \& Latorre, B. (2014). Standardized precipitation evapotranspiration index (SPEI) revisited: Parameter fitting, evapotranspiration models, tools, datasets and drought monitoring. \emph{International Journal of Climatology, 34}(10), 3001--3023. \url{https://doi.org/10.1002/joc.3887}

Behnke, R., Vavrus, S., Allstadt, A., Albright, T., Thogmartin, W. E., \& Radeloff, V. C. (2016). Evaluation of downscaled, gridded climate data for the conterminous United States. \emph{Ecological applications}, \emph{26}(5), 1338-1351.

Benali, A., Carvalho, A. C., Nunes, J. P., Carvalhais, N., \& Santos, A. (2012). Estimating air surface temperature in Portugal using MODIS LST data. \emph{Remote sensing of environment}, \emph{124}, 108-121.

Benami, E., Jin, Z., Carter, M. R., Ghosh, A., Hijmans, R. J., Hobbs, A., Kenduiywo, B., \& Lobell, D. B. (2021). Uniting remote sensing, crop modelling and economics for agricultural risk management. \emph{Nature Reviews Earth \& Environment, 2}(2): 140-59.

Borgschulte, Mark, David Molitor, and Eric Yongchen Zou. "Air pollution and the labor market: Evidence from wildfire smoke." \emph{Review of Economics and Statistics} 106, no. 6 (2024): 1558-1575.

Boryan, C., Yang, Z., Mueller, R., \& Craig, M. (2011). Monitoring US agriculture: The US Department of Agriculture, National Agricultural Statistics Service, Cropland Data Layer Program. Geocarto International, 26(5), 341--358. https://doi.org/10.1080/10106049.2011.562309

Bound, J., Brown, C., \& Mathiowetz, N. (2001). Measurement error in survey data. In \emph{Handbook of econometrics} (Vol. 5, pp. 3705-3843). Elsevier.

Bröde, P., Fiala, D., Błażejczyk, K., Holmér, I., Jendritzky, G., Kampmann, B., ... \& Havenith, G. (2012). Deriving the operational procedure for the Universal Thermal Climate Index (UTCI). \emph{International journal of biometeorology}, \emph{56}(3), 481-494.

Brown, J. F., Wardlow, B. D., Tadesse, T., Hayes, M. J., \& Reed, B. C. (2008). The Vegetation Drought Response Index (VegDRI): An integrated approach for monitoring drought-related vegetation stress. \emph{Remote Sensing of Environment, 112}(10): 3843-61.

Brown, C. F., Brumby, S. P., Guzder-Williams, B., Birch, T., Hyde, S. B., Mazzariello, J., Czerwinski, W., Pasquarella, V. J., Haertel, R., Ilyushchenko, S., Schwehr, K., Weisse, M., Stolle, F., Hanson, C., Guinan, O., Moore, R., \& Tait, A. M. (2022). Dynamic World, Near real-time global 10 m land use land cover mapping. Scientific Data, 9(1), 251. \url{https://doi.org/10.1038/s41597-022-01307-4}

Burke, M., \& Tanutama, V. (2019). Climatic constraints on aggregate economic output (No. w25779). \emph{National Bureau of Economic Research.}

Burke, M., Driscoll, A., Lobell, D. B., \& Ermon, S. (2021). Using satellite imagery to understand and promote sustainable development. \emph{Science, 371}(6535): eabe8628.

Campbell, J. B., Wynne, R. H., \& Thomas, V. A. (2022). \emph{Introduction to remote sensing} (6th ed.). Guilford Press.

Carleton, T. A., \& Hsiang, S. M. (2016). Social and economic impacts of climate. \emph{Science (New York, N.Y.)}, \emph{353}(6304), aad9837. https://doi.org/10.1126/science.aad9837

Chen, M., Shi, W., Xie, P., Silva, V. B. S., Kousky, V. E., Higgins, R. W., \& Janowiak, J. E. (2008). Assessing objective techniques for gauge-based analyses of global precipitation. \emph{Journal of Geophysical Research: Atmospheres, 113}(D4).

Chen, J. J., Mueller, V., Jia, Y., \& Tseng, S. K.-H. (2017). Validating migration responses to flooding using satellite and vital registration data. \emph{American Economic Review, 107}(5): 441-45.

Collalti, D. (2024). The economic dynamics after a flood: Evidence from satellite data. \emph{Environmental and Resource Economics, 87}(9): 2401-28.

d'Andrimont, R., Verhegghen, A., Lemoine, G., Kempeneers, P., Meroni, M., \& Van der Velde, M. (2021). From parcel to continental scale--A first European crop type map based on Sentinel-1 and LUCAS Copernicus in-situ observations. \emph{Remote Sensing of Environment, 266}: 112708.

de Bruijn, J. A., de Moel, H., Jongman, B., de Ruiter, M. C., Wagemaker, J., \& Aerts, J. C. J. H. (2019). A global database of historic and real-time flood events based on social media. \emph{Scientific Data, 6}: 311.

de Simone, L., Pizarro, E., Paredes, J., Jopia, A., Camara, G., Defourny, P., \& Bontemps, S. (2025). Quality control of training samples for agricultural statistics using Earth observation. \emph{Statistical Journal of the IAOS}: 18747655251338033.

Dee, D. P., et al. (2011). The ERA-Interim reanalysis: Configuration and performance of the data assimilation system. \emph{Quarterly Journal of the Royal Meteorological Society}, 137(656), 553--597.

Deines, J. M., Kendall, A. D., Crowley, M. A., Rapp, J., Cardille, J. A., \& Hyndman, D. W. (2019a). Mapping three decades of annual irrigation across the U.S. High Plains Aquifer using Landsat and Google Earth Engine. \emph{Remote Sensing of Environment, 233}: 111400.

Deines, J. M., Wang, S., \& Lobell, D. B. (2019b). Satellites reveal a small positive yield effect from conservation tillage across the U.S. Corn Belt. \emph{Environmental Research Letters, 14}(12): 124038.

Deines, J. M., Guan, K., Lopez, B., Zhou, Q., White, C. S., Wang, S., \& Lobell, D. B. (2023). Recent cover crop adoption is associated with small maize and soybean yield losses in the United States. \emph{Global Change Biology, 29}(3): 794-807.

Del Valle, A., Elliott, R. J. R., Strobl, E., \& Tong, M. (2018). The short-term economic impact of tropical cyclones: Satellite evidence from Guangdong Province. \emph{Economics of Disasters and Climate Change, 2}(3): 225-35.

Dell, M., Jones, B. F., \& Olken, B. A. (2014). What do we learn from the weather? The new climate-economy literature. \emph{Journal of Economic Literature, 52}(3): 740-98.

Deryugina, T., et al. (2019). The mortality and medical costs of air pollution: Evidence from changes in wind direction. \emph{American Economic Review}, 109(12), 4178--4219.

Diao, C. (2020). Remote sensing phenological monitoring framework to characterize corn and soybean physiological growing stages. Remote Sensing of Environment, 248, 111960. \url{https://doi.org/10.1016/j.rse.2020.111960}

Dinku, T., Funk, C., Peterson, P., Maidment, R., Tadesse, T., Gadain, H., \& Ceccato, P. (2018). Validation of the CHIRPS satellite rainfall estimates over eastern Africa. \emph{Quarterly Journal of the Royal Meteorological Society}, \emph{144}, 292-312.

Deryng, D., Conway, D., Ramankutty, N., Price, J., \& Warren, R. (2014). Global crop yield response to extreme heat stress under multiple climate change futures. \emph{Environmental Research Letters, 9}: 034011.

Di Baldassarre, G., Viglione, A., Carr, G., Kuil, L., Yan, K., Brandimarte, L., \& Blöschl, G. (2015). Debates---Perspectives on socio-hydrology: Capturing feedbacks between physical and social processes. \emph{Water Resources Research, 51}: 4770-81.

Di Falco, S., Berck, P., Bezabih, M., \& Köhlin, G. (2019). Rain and impatience: Evidence from rural Ethiopia. \emph{Journal of Economic Behavior \& Organization, 160}: 40-51.

Di Falco, S., Feri, F., Pin, P., \& Vollenweider, X. (2018). Ties that bind: Network redistributive pressure and economic decisions in village economies. \emph{Journal of Development Economics, 131}: 123-31.

Diao, C. (2020). Remote sensing phenological monitoring framework to characterize corn and soybean physiological growing stages. \emph{Remote Sensing of Environment, 248}: 111960.

Doorenbos, J., \& Kassam, A. H. (1979). \emph{Yield Response to Water}. FAO Irrigation and Drainage Paper No. 33. Food and Agriculture Organization, Rome.

Dunne, J. P., Stouffer, R. J., \& John, J. G. (2013). Reductions in labour capacity from heat stress under climate warming. \emph{Nature Climate Change}, \emph{3}(6), 563-566.

ECMWF. (2025). \emph{Scientific and technical support for AgERA5 ending from 31-12-2025} {[}Forum post{]}. ECMWF Forum. \url{https://forum.ecmwf.int/t/scientific-and-technical-support-for-agera5-ending-from-31-12-2025/14468}

Enenkel, M., Osgood, D., Anderson, M., Powell, B., McCarty, J., Neigh, C., Carroll, M., Wooten, M., Husak, G., Hain, C., \& Brown, M. (2019). Exploiting the convergence of evidence in satellite data for advanced weather index insurance design. \emph{Weather, Climate, and Society, 11}: 65-93.

Entekhabi, D., Njoku, E. G., O\textquotesingle neill, P. E., Kellogg, K. H., Crow, W. T., Edelstein, W. N., ... \& Van Zyl, J. (2010). The soil moisture active passive (SMAP) mission. \emph{Proceedings of the IEEE}, \emph{98}(5), 704-716.

Estes, L., Chen, P., Debats, S., Evans, T., Ferreira, S., Kuemmerle, T., Ragazzo, G., Sheffield, J., Wolf, A., Wood, E., \& Caylor, K. (2018). A large-area, spatially continuous assessment of land cover map error and its impact on downstream analyses. \emph{Global Change Biology, 24}(1): 322-37.

Estes, L. D., Ye, S., Song, L., Luo, B., Eastman, J. R., Meng, Z., Zhang, Q., McRitchie, J., Debats, S., Muhando, J., Amukoa, E., Kaloo, P., Makuru, B., Mbatia, O., Muasa, M., Mucha, A., Mugami, F., Muinde, P., Mwawaza, S., Ochieng, J., Oduol, C., Oduor, C., Wanjiku, R., Wanyoike, A., Avery, R., \& Caylor, K. K. (2022). High resolution, annual maps of field boundaries for smallholder-dominated croplands at national scales. Frontiers in Artificial Intelligence, 4: 744863.

FAO. (2020). WaPOR Database Methodology: \emph{V}ersion 2 release.~ https://doi.org/10.4060/ca9894en

Fischer, E. M., \& Knutti, R. (2015). Anthropogenic contribution to global occurrence of heavy-precipitation and high-temperature extremes. \emph{Nature climate change}, \emph{5}(6), 560-564.

Fisher, J. B., et al. (2017). The future of evapotranspiration: Global requirements for ecosystem functioning, carbon and climate feedbacks, agricultural management, and water resources. \emph{Water Resources Research}, 53(4), 2618--2626.~

Fowlie, M., Rubin, E., \& Walker, R. (2019). Bringing satellite-based air quality estimates down to earth. \emph{AEA Papers and Proceedings, 109}: 283-88.

Fritz, S., See, L., McCallum, I., You, L., Bun, A., Moltchanova, E., Duerauer, M., Albrecht, F., Schill, C., Perger, C., Havlik, P., Mosnier, A., Thornton, P., Wood-Sichra, U., Herrero, M., Becker-Reshef, I., Justice, C., Hansen, M., Gong, P., Abdel Aziz, S., Cipriani, A., Cumani, R., Cecchi, G., Conchedda, G., Ferreira, S., Gomez, A., Haffani, M., Kayitakire, F., Malanding, J., Mueller, R., Newby, T., Nonguierma, A., Olusegun, A., Ortner, S., Rajak, D. R., Rocha, J., Schepaschenko, D., Schepaschenko, M., Terekhov, A., Tiangwa, A., Vancutsem, C., Vintrou, E., Wenbin, W., van der Velde, M., Dunwoody, A., Kraxner, F., \& Obersteiner, M. (2015). Mapping global cropland and field size. \emph{Global Change Biology, 21}: 1980-92.

Fritz, S., See, L., Bayas, J. C. L., Waldner, F., Jacques, D., Becker-Reshef, I., Whitcraft, A., Baruth, B., Bonifacio, R., Crutchfield, J., Rembold, F., Rojas, O., Schucknecht, A., Van der Velde, M., Verdin, J., Wu, B., Yan, N., You, L., Gilliams, S., Mücher, S., Tetrault, R., Moorthy, I., \& McCallum, I. (2019). A comparison of global agricultural monitoring systems and current gaps. \emph{Agricultural Systems, 168}: 258-72.

Funk, C., Peterson, P., Landsfeld, M., Pedreros, D., Verdin, J., Shukla, S., Husak, G., Rowland, J., Harrison, L., Hoell, A., \& Michaelsen, J. (2015). The climate hazards infrared precipitation with stations---A new environmental record for monitoring extremes. \emph{Scientific Data}, \emph{2}, 150066

Funk, C., Peterson, P., Peterson, S., Shukla, S., Davenport, F., Michaelsen, J., ... \& Mata, N. (2019). A high-resolution 1983--2016 T max climate data record based on infrared temperatures and stations by the Climate Hazard Center. \emph{Journal of Climate}, \emph{32}(17), 5639-5658.

García, A., \& Heilmayr, R. (2024). Impact evaluation with nonrepeatable outcomes: The case of forest conservation. \emph{Journal of Environmental Economics and Management, 124}, Article 102971. \url{https://doi.org/10.1016/j.jeem.2024.102971}

Garg, T., Jagnani, M., Kresch, E., \& Chindo, M. (2020). Temperature and human capital in India. \emph{Journal of the Association of Environmental and Resource Economists, 7}(6): 1179-1217.

Gelaro, R., McCarty, W., Suarez, M. J., Todling, R., Molod, A., Takacs, L., Randles, C. A., Darmenov, A., Bosilovich, M. G., Reichle, R., Wargan, K., Coy, L., Cullather, R., Draper, C., Akella, S., Buchard, V., Conaty, A., da Silva, A. M., Gu, W., Kim, G.-K., Koster, R., Lucchesi, R., Merkova, D., Nielsen, J. E., Paryka, G., Pawson, S., Putnam, W., Rienecker, M., Schubert, S. D., Sienkiewicz, M., \& Zhao, B. (2017). The Modern-Era Retrospective Analysis for Research and Applications, Version 2 (MERRA-2). \emph{Journal of Climate, 30}(14): 10.1175/JCLIM-D-16-0758.1.

Giezendanner, J., Mukherjee, R., Purri, M., Thomas, M., Mauerman, M., Islam, A., \& Tellman, B. (2023). Inferring the past: A combined CNN-LSTM deep learning framework to fuse satellites for historical inundation mapping. In \emph{Proceedings of the IEEE/CVF Conference on Computer Vision and Pattern Recognition} (pp. 2155-65).

Gitelson, A. A., Viña, A., Arkebauer, T. J., Rundquist, D. C., Keydan, G., \& Leavitt, B. (2003). Remote estimation of leaf area index and green leaf biomass in maize canopies. \emph{Geophysical Research Letters, 30}(5).

Good, E. J., Ghent, D. J., Bulgin, C. E., \& Remedios, J. J. (2017). A spatiotemporal analysis of the relationship between near‐surface air temperature and satellite land surface temperatures using 17 years of data from the ATSR series. \emph{Journal of Geophysical Research: Atmospheres}, \emph{122}(17), 9185-9210.

Gordon, M., Stone, E., Ayers, M., \& Sanford, L. (2026). \emph{Debiasing estimates of global forest cover loss} {[}Working paper{]}. \url{https://mdgordo.github.io/personalwebsite/Adversarial_Debiasing_for_Unbiased_Parameter_Recovery__P_P_.pdf}

Graff Zivin, J., \& Neidell, M. (2014). Temperature and the allocation of time: Implications for climate change. \emph{Journal of Labor Economics}, \emph{32}(1), 1-26.

Greßer, C., Meierrieks, D., \& Stadelmann, D. (2021). The link between regional temperature and regional incomes: Econometric evidence with sub-national data. \emph{Economic Policy, 36}(107): 523-50.

Grimaldi, S., Schumann, G. P., Shokri, A., Walker, J. P., \& Pauwels, V. R. N. (2019). Challenges, opportunities, and pitfalls for global coupled hydrologic‐hydraulic modeling of floods. \emph{Water Resources Research}, \emph{55}(7), 5277-5300.

Grossiord, C., Buckley, T. N., Cernusak, L. A., Novick, K. A., Poulter, B., Siegwolf, R. T., ... \& McDowell, N. G. (2020). Plant responses to rising vapor pressure deficit. \emph{New phytologist}, \emph{226}(6), 1550-1566.

Guiteras, R., Jina, A., \& Mobarak, A. M. (2015). Satellites, self-reports, and submersion: Exposure to floods in Bangladesh. \emph{American Economic Review, 105}(5): 232-36.

Harrigan, S., Zsoter, E., Alfieri, L., Prudhomme, C., Salamon, P., Wetterhall, F., ... \& Pappenberger, F. (2020). GloFAS-ERA5 operational global river discharge reanalysis 1979--present. \emph{Earth System Science Data Discussions}, \emph{2020}, 1-23.

Hatfield, J.L. and Prueger, J.H., 2015. Temperature extremes: Effect on plant growth and development. \emph{Weather and climate extremes}, \emph{10}, pp.4-10.

Hausman, J. (2001). Mismeasured variables in econometric analysis: Problems from the right and problems from the left. \emph{Journal of Economic Perspectives, 15}(4): 57-77.

Hersbach, H., Bell, B., Berrisford, P., Hirahara, S., Horanyi, A., Munoz-Sabater, J., Nicolas, J., Peavey, C., Radu, R., Schepers, D., Simmons, A., Soci, C., Abdalla, S., Abellan, X., Balsamo, G., Bechtold, P., Biavati, G., Bidlot, J., Bonavita, M., Chiara, G., Dalgren, P., Dee, D., Diamantakias, M., Dragani, R., Fleming, J., Forbes, R., Fuentes, M., Geer, A., Haimberger, L., Healy, S., Hogan, R. J., Holm, E., Janiskova, M., Keeley, S., Laloyaux, P., Lopez, P., Lupu, C., Radnoti, G., Rosnay, P., Rozum, I., Vamborg, F., Villaume, S., \& Thepaut, J.-N. (2020). The ERA5 global reanalysis. \emph{Quarterly Journal of the Royal Meteorological Society, 146}(730): 1999-2049.

Huete, A. R. (1988). A soil-adjusted vegetation index (SAVI). \emph{Remote Sensing of Environment, 25}(3): 295-309.

Huete, A., Didan, K., Miura, T., Rodriguez, E. P., Gao, X., \& Ferreira, L. G. (2002). Overview of the radiometric and biophysical performance of the MODIS vegetation indices. \emph{Remote Sensing of Environment, 83}(1-2): 195-213.

Huffman, G. J., Bolvin, D. T., Braithwaite, D., Hsu, K. L., Joyce, R. J., Kidd, C., ... \& Xie, P. (2020). Integrated multi-satellite retrievals for the global precipitation measurement (GPM) mission (IMERG). In \emph{Satellite precipitation measurement: Volume 1} (pp. 343-353). Cham: Springer International Publishing.

~Hulley, G. C., Malakar, N. K., Islam, T., \& Freepartner, R. J. (2017). NASA\textquotesingle s MODIS and VIIRS land surface temperature and emissivity products: A long-term and consistent earth system data record. \emph{IEEE Journal of Selected Topics in Applied Earth Observations and Remote Sensing}, \emph{11}(2), 522-535.

Immerzeel, W. W., Van Beek, L. P., \& Bierkens, M. F. (2010). Climate change will affect the Asian water towers. Science, 328(5984), 1382-1385.

Jack, K., \& Walker, K. (2024). Integrating remote sensing and randomized controlled trials. Working paper. \url{https://kelseyjack.bren.ucsb.edu/research/RS-RCT-guidelines}

Jain, M. (2020). The benefits and pitfalls of using satellite data for causal inference. \emph{Review of Environmental Economics and Policy, 14}(1): 157-69.

Jiménez, P. A., González-Rouco, J. F., Navarro, J., Montávez, J. P., \& García-Bustamante, E. (2010). Quality assurance of surface wind observations from automated weather stations. \emph{Journal of Atmospheric and Oceanic Technology}, \emph{27}(7), 1101-1122.

Jongman, B., Wagemaker, J., Romero, B. R., \& De Perez, E. C. (2015). Early flood detection for rapid humanitarian response: Harnessing near real-time satellite and Twitter signals. \emph{ISPRS International Journal of Geo-Information, 4}(4): 2246-66.

Josephson, A., Michler, J. D., Kilic, T., \& Murray, S. (2025). The mismeasure of weather: Using remotely sensed earth observation data in economic context.

Justice, C. O., Vermote, E., Townshend, J. R., Defries, R., Roy, D. P., Hall, D. K., ... \& Barnsley, M. J. (1998). The Moderate Resolution Imaging Spectroradiometer (MODIS): Land remote sensing for global change research. \emph{IEEE transactions on geoscience and remote sensing}, \emph{36}(4), 1228-1249.

Kalnay, E., Kanamitsu, M., Kistler, R., Collins, W., Deaven, D., Gandin, L., ... \& Joseph, D. (1996). \emph{The ncep/ncar 40-year reanalysis project, B. Am. Meteorol. Soc., 77, 437--471}.

Karlsson, J. (2021). Temperature and exports: Evidence from the United States. \emph{Environmental and Resource Economics, 80}(2): 311-37.

Kaufman, Y. J., \& Tanre, D. (1992). Atmospherically resistant vegetation index (ARVI) for EOS-MODIS. \emph{IEEE transactions on Geoscience and Remote Sensing}, \emph{30}(2), 261-270.

Kerner, H., Nakalembe, C., Yang, A., Zvonkov, I., McWeeny, R., Tseng, G., \& Becker-Reshef, I. (2024). How accurate are existing land cover maps for agriculture in Sub-Saharan Africa? Scientific Data, 11(1), 486. https://doi.org/10.1038/s41597-024-03306-z

Kerner, H., Chaudhari, S., Ghosh, A., Robinson, C., Ahmad, A., Choi, E., Jacobs, N., Holmes, C., Mohr, M., Dodhia, R., Lavista Ferres, J. M., \& Marcus, J. (2025). Fields of the world: A machine learning benchmark dataset for global agricultural field boundary segmentation. \emph{Proceedings of the AAAI Conference on Artificial Intelligence}, 39(27): 28151-28159

Kjellstrom, T., Holmer, I., \& Lemke, B. (2009). Workplace heat stress, health and productivity--an increasing challenge for low and middle-income countries during climate change. \emph{Global health action}, \emph{2}(1), 2047.

Kocornik-Mina, A., McDermott, T. K., Michaels, G., \& Rauch, F. (2020). Flooded cities. \emph{American Economic Journal: Applied Economics}, \emph{12}(2), 35-66.

Kosaka, Y., Kobayashi, S., Harada, Y., Kobayashi, C., Naoe, H., Yoshimoto, K., Harada, M., Goto, N., Chiba, J., Miyaoka, K., et al. (2024). The JRA-3Q Reanalysis. \emph{Journal of the Meteorological Society of Japan, 102}(1): 49-109.

Labus, M. P., Nielsen, G. A., Lawrence, R. L., Engel, R., \& Long, D. S. (2002). Wheat yield estimates using multi-temporal NDVI satellite imagery. \emph{International Journal of Remote Sensing, 23}(20): 4169-80.

Langlois, B. K., Magnuson, A., Griffin, T., Coughlan de Perez, E., Naumova, E., \& Koch, M. (2024). A framework and pilot study for assessing usability of flood data portals for interdisciplinary research. \emph{PLOS Climate}, \emph{3}(11), e0000511.

Lark, T. J., Schelly, I. H., \& Gibbs, H. K. (2021). Accuracy, Bias, and Improvements in Mapping Crops and Cropland across the United States Using the USDA Cropland Data Layer. Remote Sensing, 13(5), 968. https://doi.org/10.3390/rs13050968

Liu, W., Gopal, S., \& Woodcock, C. E. (2004). Uncertainty and Confidence in Land Cover Classification Using a Hybrid Classifier Approach. Photogrammetric Engineering \& Remote Sensing, 70(8), 963--971. https://doi.org/10.14358/PERS.70.8.963

Liu, T., Mickley, L. J., Marlier, M. E., DeFries, R. S., Khan, M. F., Latif, M. T., \& Karambelas, A. (2020). Diagnosing spatial biases and uncertainties in global fire emissions inventories: Indonesia as regional case study. \emph{Remote Sensing of Environment, 237}: 111557.

Lobell, D. B., \& Field, C. B. (2007). Global scale climate--crop yield relationships and the impacts of recent warming. \emph{Environmental Research Letters, 2}(1): 014002.

Lobell, D. B., Schlenker, W., \& Costa-Roberts, J. (2011). Climate trends and global crop production since 1980. \emph{Science, 333}(6042): 616-20.

Lobell, D. B., Ortiz-Monasterio, J. I., Sibley, A. M., \& Sohu, V. S. (2013). Satellite detection of earlier wheat sowing in India and implications for yield trends. Agricultural Systems, 115, 137--143. https://doi.org/10.1016/j.agsy.2012.09.003

Lobell, D. B. (2013). The use of satellite data for crop yield gap analysis. \emph{Field Crops Research, 143}: 56-64.

Lobell, D. B., Thau, D., Seifert, C., Engle, E., \& Little, B. (2015). A scalable satellite-based crop yield mapper. \emph{Remote Sensing of Environment, 164}: 324-33.

Lobell, D. B., Azzari, G., Burke, M., Gourlay, S., Jin, Z., Kilic, T., \& Murray, S. (2020). Eyes in the sky, boots on the ground: Assessing satellite- and ground-based approaches to crop yield measurement and analysis. \emph{American Journal of Agricultural Economics, 102}(1): 202-19.

Loeb, N. G., Doelling, D. R., Wang, H., Su, W., Nguyen, C., Corbett, J. G., ... \& Kato, S. (2018). Clouds and the earth's radiant energy system (CERES) energy balanced and filled (EBAF) top-of-atmosphere (TOA) edition-4.0 data product. \emph{Journal of climate}, \emph{31}(2), 895-918.

Loeb, N. G., Kato, S., Rose, F. G., Doelling, D. R., Rutan, D. A., Su, W., \& Smith, W. L. (2018). Toward a consistent definition between satellite and model clear-sky radiative fluxes. \emph{Journal of Climate, 31}: 231-50.

Lorenz, C., \& Kunstmann, H. (2012). The hydrological cycle in three state-of-the-art reanalyses: Intercomparison and performance analysis. \emph{Journal of Hydrometeorology}, 13(5), 1397--1420.

Maleki, R., Wu, F., Oubara, A., Fathollahi, L., \& Yang, G. (2024). Refinement of Cropland Data Layer with Effective Confidence Layer Interval and Image Filtering. Agriculture, 14(8), 1285. https://doi.org/10.3390/agriculture14081285

Mannaerts, C. M., Blatchford, M. L., Njuki, S. M., Zeng, Y., Nouri, H., \& Maathuis, B. H. P. (2020). \emph{WaPOR quality assessment: Technical report on the data quality of the WaPOR FAO database version 2}. Food and Agriculture Organization of the United Nations.

Mayorga, J., Villacis, A. H., \& Mishra, A. K. (2024). Farm-level agricultural productivity and adaptation to extreme heat. \emph{American Journal of Agricultural Economics.} \url{https://doi.org/10.1111/ajae.12509}

McNally, A., Husak, G. J., Brown, M., Carroll, M., Funk, C., Yatheendradas, S., \& Verdin, J. P. (2015). Calculating crop water requirement satisfaction in the West Africa Sahel with remotely sensed soil moisture. \emph{Journal of Hydrometeorology, 16}(1): 295--305.

Menne, M. J., Durre, I., Vose, R. S., Gleason, B. E., \& Houston, T. G. (2012). An overview of the Global Historical Climatology Network-Daily Database. \emph{Journal of Atmospheric and Oceanic Technology, 29}: 897-910.

Melton, F. S., Huntington, J., Grimm, R., Herring, J., Hall, M., Rollison, D., ... \& Anderson, R. G. (2022). OpenET: Filling a critical data gap in water management for the western United States. \emph{JAWRA Journal of the American Water Resources Association}, \emph{58}(6), 971-994.

Michler, J. D., Al Rafi, D. A., Giezendanner, J., Josephson, A., Pede, V. O., \& Tellman, E. (2026). Impact evaluations in data-scarce environments: The case of stress-tolerant rice varieties in Bangladesh. \emph{Journal of Development Economics}, 103648.

Mildrexler, D. J., Zhao, M., \& Running, S. W. (2011). A global comparison between station air temperatures and MODIS land surface temperatures reveals the cooling role of forests. \emph{Journal of Geophysical Research: Biogeosciences}, \emph{116}(G3).

Mishra, A. K., \& Singh, V. P. (2010). A review of drought concepts. \emph{Journal of Hydrology, 391}(1-2): 202-16.

Mishra, A., Mukherjee, S., Merz, B., Singh, V. P., Wright, D. B., Villarini, G., ... \& Stedinger, J. R. (2022). An overview of flood concepts, challenges, and future directions. \emph{Journal of hydrologic engineering}, \emph{27}(6), 03122001.

Mladenova, I. E., Bolten, J. D., Crow, W., Sazib, N., \& Reynolds, C. (2020). Agricultural drought monitoring via the assimilation of SMAP soil moisture retrievals into a global soil water balance model. \emph{Frontiers in Big Data}, \emph{3}, 10.

Moon, M., Zhang, X., Henebry, G. M., Liu, L., Gray, J. M., Melaas, E. K., \& Friedl, M. A. (2019). Long-term continuity in land surface phenology measurements: A comparative assessment of the MODIS land cover dynamics and VIIRS land surface phenology products. Remote Sensing of Environment, 226, 74--92. \url{https://doi.org/10.1016/j.rse.2019.03.034}~

Moore, F. C., \& Obradovich, N. (2020). Using remarkability to define coastal flooding thresholds. \emph{Nature Communications, 11}(1): 530.

Müller, R., \& Pfeifroth, U. (2022). Remote sensing of solar surface radiation--a reflection of concepts, applications and input data based on experience with the effective cloud albedo. \emph{Atmospheric Measurement Techniques}, \emph{15}(5), 1537-1561.

Nabil, M., Zhang, M., Bofana, J., Wu, B., Stein, A., Dong, T., ... \& Shang, J. (2020). Assessing factors impacting the spatial discrepancy of remote sensing based cropland products: A case study in Africa. \emph{International Journal of Applied Earth Observation and Geoinformation}, \emph{85}, 102010.

NASA Earthdata. (2025, May 2). Transition from MODIS to VIIRS. NASA. \url{https://www.earthdata.nasa.gov/data/alerts-outages/transition-from-modis-viirs}

NASA Earthdata. (n.d.). Data Processing Levels. NASA. Retrieved February 2, 2026, from \url{https://www.earthdata.nasa.gov/learn/earth-observation-data-basics/data-processing-levels}\ul{~}

Olofsson, P., Foody, G. M., Herold, M., Stehman, S. V., Woodcock, C. E., \& Wulder, M. A. (2014). Good practices for estimating area and assessing accuracy of land change. \emph{Remote Sensing of Environment, 148}: 42-57.

Ortiz-Bobea, A., Ault, T. R., Carrillo, C. M., Chambers, R. G., \& Lobell, D. B. (2021). Large increases in public R\&D investment are needed to avoid declines of U.S. agricultural productivity. \emph{PNAS, 122}(11): e2411010122.

Ortiz-Bobea, A., Chambers, R. G., He, Y., \& Lobell, D. B. (2025). Anthropogenic climate change has slowed global agricultural productivity growth. \emph{Nature Climate Change, 11}: 306-12.

Palmer, W. (1965). Meteorological drought. Research Paper no. 45, U.S. Department of Commerce Weather Bureau. Available online by the NOAA National Climatic Data Center at \url{http://www.ncdc.noaa.gov/temp-and-precip/drought/docs/palmer.pdf}

Parker, W.S., 2016. Reanalyses and observations: What's the difference?. \emph{Bulletin of the American Meteorological Society}, \emph{97}(9), pp.1565-1572.

Patel, D. (2023). Floods. \emph{Unpublished dissertation.} Available at SSRN: \url{https://ssrn.com/abstract=4636828}

Pope III, C. A., \& Dockery, D. W. (2006). Health effects of fine particulate air pollution: lines that connect. \emph{Journal of the air \& waste management association}, \emph{56}(6), 709-742.

Proctor, J., Carleton, T., \& Sum, S. (2023). Parameter recovery using remotely sensed variables (No. w30861). \emph{National Bureau of Economic Research.}

Ratledge, N., Cadamuro, G., De La Cuesta, B., Stigler, M., \& Burke, M. (2022). Using machine learning to assess the livelihood impact of electricity access. \emph{Nature, 611}(7936): 491-95.

Raymond, C., Matthews, T., \& Horton, R. M. (2020). The emergence of heat and humidity too severe for human tolerance. \emph{Science Advances}, \emph{6}(19), eaaw1838.

Remelgado, R., Conrad, C., \& Meyer, C. (2023). Spatiotemporal inconsistencies in Landsat satellite observations bias environmental-change analyses and monitoring.

Richaud, B., Anthonj, J., Larsen, O., Enenkel, M., \& Plevin, J. L. (2024). NextGen Drought Index Dashboard -- Designing and piloting a new satellite data platform to strengthen drought risk financing in the Horn of Africa. EGU General Assembly 2024, Vienna, Austria, 14--19 Apr 2024, EGU24-22325.

Rockström, J., et al. (2007). Managing water in rainfed agriculture. In \emph{Water for Food, Water for Life: A Comprehensive Assessment of Water Management in Agriculture}. Earthscan/IWMI.

Rohde, R. A., \& Hausfather, Z. (2020). The Berkeley Earth land/ocean temperature record. \emph{Earth System Science Data}, \emph{12}(4), 3469-3479.

Rouse, J. W., Haas, R. H., Deering, D. W., Schell, J. A., \& Harlan, J. C. (1974). Monitoring the vernal advancement and retrogradation (green wave effect) of natural vegetation (No. E75-10354).

Schauberger, B., Archontoulis, S., Arneth, A., et al. (2017). Consistent negative response of U.S. crops to high temperatures in observations and crop models. \emph{Nature Communications, 8}: 13931.

Schennach, S. M. (2016). Recent advances in the measurement error literature. \emph{Annual Review of Economics, 8}: 341-77.

Schultz, K. A., \& Mankin, J. S. (2019). Is temperature exogenous? The impact of civil conflict on the instrumental climate record in Sub‐Saharan Africa. \emph{American Journal of Political Science, 63}(4): 723-39.

Seager, R., et al. (2018). Whither the 100th meridian? \emph{Earth Interactions}, 22(5), 1--23.

Seifert, C. A., Azzari, G., \& Lobell, D. B. (2018). Satellite detection of cover crops and their effects on crop yield in the Midwestern United States. \emph{Environmental Research Letters, 13}.

Senay, G. B., \& Verdin, J. (2003). Characterization of yield reduction in Ethiopia using a GIS-based crop water balance model. \emph{Canadian Journal of Remote Sensing}, \emph{29}(6), 687-692.

Senay, G. B., Kagone, S., \& Velpuri, N. M. (2020). Operational global actual evapotranspiration: Development, evaluation, and dissemination. \emph{Sensors}, \emph{20}(7), 1915.

Shah, M., \& Steinberg, B. M. (2017). Drought of opportunities: Contemporaneous and long-term impacts of rainfall shocks on human capital. \emph{Journal of Political Economy, 125}(2): 527-61.

Shen, Y., Zhang, X., \& Yang, Z. (2022). Mapping corn and soybean phenometrics at field scales over the United States Corn Belt by fusing time series of Landsat 8 and Sentinel-2 data with VIIRS data. \emph{ISPRS Journal of Photogrammetry and Remote Sensing, 186}: 55-69.

Siebert, S., et al. (2010). Groundwater use for irrigation --- a global inventory. \emph{Hydrology and Earth System Sciences}, 14, 1863--1880

Simó, G., Martínez-Villagrasa, D., Jiménez, M. A., Caselles, V., \& Cuxart, J. (2019). Impact of the surface--atmosphere variables on the relation between air and land surface temperatures. \emph{Pure and Applied Geophysics, 175}: 3939-53.

Smith, H. D., Dubeux, J. C., Zare, A., \& Wilson, C. H. (2023). Assessing transferability of remote sensing pasture estimates using multiple machine learning algorithms and evaluation structures. \emph{Remote Sensing, 15}(11): 2940.

Singh, T. P. (2022). Beyond the haze: Air pollution and student absenteeism---Evidence from India. \emph{OSF}. \url{https://doi.org/10.31219/osf.io/pcva2}

Sohnesen, T. P. (2020). Two sides to same drought: Measurement and impact of Ethiopia's 2015 historical drought. \emph{Economics of Disasters and Climate Change, 4}(1): 83-101.

Steadman, R. G. (1979). The assessment of sultriness. Part I: A temperature-humidity index based on human physiology and clothing science. \emph{Journal of Applied Meteorology}, 18(7), 861--873.~

Steduto, P., et al. (2012). \emph{Crop Yield Response to Water}. FAO Irrigation and Drainage Paper 66. Food and Agriculture Organization, Rome.

Stephens, G.L. and Kummerow, C.D., 2007. The remote sensing of clouds and precipitation from space: A review. \emph{Journal of the Atmospheric Sciences}, \emph{64}(11), pp.3742-3765.

Stull, R. (2011). Wet-bulb temperature from relative humidity and air temperature. \emph{Journal of applied meteorology and climatology}, \emph{50}(11), 2267-2269.

Sun, Q., Miao, C., Duan, Q., Ashouri, H., Sorooshian, S., \& Hsu, K. L. (2018). A review of global precipitation data sets: Data sources, estimation, and intercomparisons. \emph{Reviews of geophysics}, \emph{56}(1), 79-107.

Susskind, J., Blaisdell, J., \& Iredell, L. (2014). Improved methodology for surface and atmospheric soundings, error estimates, and quality-control procedures: the AIRS Science Team Version-6 retrieval algorithm. \emph{Journal of Applied Remote Sensing, 8}(1): 084994.

Tellman, B., Sullivan, J. A., Kuhn, C., Kettner, A. J., Doyle, C. S., Brakenridge, G. R., ... \& Slayback, D. A. (2021). Satellite imaging reveals increased proportion of population exposed to floods. \emph{Nature}, \emph{596}(7870), 80-86.Chen, J.J., Mueller, V., Jia, Y., \& Tseng, S.K.-H. (2017). Validating migration responses to flooding using satellite and vital registration data. \emph{American Economic Review}, 107(5): 441--445.

Tian, Y., Peters-Lidard, C. D., Eylander, J. B., Joyce, R. J., Huffman, G. J., Adler, R. F., Hsu, K., Turk, F. J., Garcia, M., and Zeng, J. (2009). Component analysis of errors in satellite-based precipitation estimates. \emph{Journal of Geophysical Research}, 114, D24101. \url{https://doi.org/10.1029/2009JD011949}

Tian, Y., \& Peters‐Lidard, C. D. (2010). A global map of uncertainties in satellite‐based precipitation measurements. \emph{Geophysical Research Letters}, \emph{37}(24).

Townshend, J. R., Justice, C. O., Gurney, C., \& McManus, J. (1992). The impact of misregistration on change detection. \emph{IEEE Transactions on Geoscience and remote sensing}, \emph{30}(5), 1054-1060.

Tuholske, C., Caylor, K., Funk, C., Verdin, A., Sweeney, S., Grace, K., Peterson, P., \& Evans, T. (2021). Global urban population exposure to extreme heat. \emph{Proceedings of the National Academy of Sciences, 118}(41): e2024792118.

van Donkelaar, A., Hammer, M. S., Bindle, L., et al. (2021). Monthly global estimates of fine particulate matter and their uncertainty. \emph{Environmental Science \& Technology, 55}: 15287-300.

Van Tricht, K., Degerickx, J., Gilliams, S., Zanaga, D., Battude, M., Grosu, A., Djomeni, M., Defourny, P., Verger, A., Dorigo, W., Estel, S., Tychon, B., Martinez-Sanchez, L., Wu, J., Szantoi, Z., \& the WorldCereal Consortium. (2023). WorldCereal: A dynamic open-source system for global-scale, seasonal, and reproducible crop and irrigation mapping. \emph{Earth System Science Data}, 15(12): 5491-5515.

Venter, Z. S., Barton, D. N., Chakraborty, T., Simensen, T., \& Singh, G. (2022). Global 10 m land use land cover datasets: A comparison of dynamic world, world cover and esri land cover. \emph{Remote Sensing}, \emph{14}(16), 4101.

Verdin, J., \& Klaver, R. (2002). Grid-cell-based crop water accounting for the famine early warning system. \emph{Hydrological Processes, 16}: 1617-30.

Verdin, A., Funk, C., Peterson, P., Landsfeld, M., Tuholske, C., \& Grace, K. (2020). Development and validation of the CHIRTS-daily quasi-global high-resolution daily temperature data set. \emph{Scientific Data, 7}: 303.

Vermote, E. F., El Saleous, N., Justice, C. O., Kaufman, Y. J., Privette, J. L., Remer, L., ... \& Tanre, D. (1997). Atmospheric correction of visible to middle‐infrared EOS‐MODIS data over land surfaces: Background, operational algorithm and validation. \emph{Journal of Geophysical Research: Atmospheres}, \emph{102}(D14), 17131-17141.

Vicente-Serrano, S. M., Beguería, S., \& López-Moreno, J. I. (2010). A multiscalar drought index sensitive to global warming: The standardized precipitation evapotranspiration index. \emph{Journal of Climate, 23}(7): 1696--1718.

Wagner, W., Hahn, S., Kidd, R., Melzer, T., Bartalis, Z., Hasenauer, S., ... \& Rubel, F. (2013). The ASCAT soil moisture product: A review of its specifications, validation results, and emerging applications. \emph{Meteorologische Zeitschrift}.

Wang, X., Liu, Z., \& Chen, H. (2022). Investigating flood impact on crop production under a comprehensive and spatially explicit risk evaluation framework. \emph{Agriculture, 12}: 484.

Weiss, M., Jacob, F., \& Duveiller, G. (2020). Remote sensing for agricultural applications: A meta-review. \emph{Remote Sensing of Environment, 236}: 111402.

Wilson, A. J., Bressler, R. D., Ivanovich, C., Tuholske, C., Raymond, C., Horton, R. M., Sobel, A., Kinney, P., Cavazos, T., \& Shrader, J. G. (2024). Heat disproportionately kills young people: Evidence from wet-bulb temperature in Mexico. \emph{Science Advances, 10}(49): eadq3367.

World Meteorological Organization (WMO). (n.d.). Climate. \url{https://wmo.int/topics/climate}

World Meteorological Organization. (2025). Guide to instruments and methods of observation: Volume I---Measurement of meteorological variables (2024 ed., WMO-No. 8). WMO. \url{https://library.wmo.int/records/item/68695-guide-to-instruments-and-methods-of-observation}

Wuepper, D., Oluoch, W. A., \& Hadi, H. (2025). Satellite data in agricultural and environmental economics: Theory and practice. \emph{Agricultural Economics, 56}(3): 493--511.

Xu, P., Tsendbazar, N. E., Herold, M., De Bruin, S., Koopmans, M., Birch, T., ... \& Zanaga, D. (2024). Comparative validation of recent 10 m-resolution global land cover maps. \emph{Remote Sensing of Environment}, \emph{311}, 114316.

Yuan, W., Zheng, Y., Piao, S., Ciais, P., Lombardozzi, D., Wang, Y., ... \& Yang, S. (2019). Increased atmospheric vapor pressure deficit reduces global vegetation growth. \emph{Science advances}, \emph{5}(8), eaax1396.

Zhang, T., Zhou, Y., Zhao, K., Zhu, Z., Chen, G., Hu, J., \& Wang, L. (2022). A global dataset of daily maximum and minimum near-surface air temperature at 1 km resolution over land (2003--2020). \emph{Earth System Science Data, 14}: 5637--49.

\clearpage

\end{document}